\documentclass[12pt] {article}
\usepackage{psfrag}
\usepackage{graphicx}
\usepackage{latexsym,amsfonts}
\usepackage{amssymb}
\begin{document}

\setcounter{page}{1}
\def\theequation{\arabic{section}.\arabic{equation}}
\def\theequation{\thesection.\arabic{equation}}
\setcounter{section}{0}

\title{On isospin--breaking corrections\\ to the energy level
  displacement of the ground state of kaonic hydrogen}

\author{A. N. Ivanov\,\thanks{E--mail: ivanov@kph.tuwien.ac.at, Tel.:
+43--1--58801--14261, Fax: +43--1--58801--14299}~\thanks{Permanent
Address: State Polytechnic University, Department of Nuclear Physics,
195251 St. Petersburg, Russian Federation}\,,
M. Cargnelli\,\thanks{E--mail: michael.cargnelli@oeaw.ac.at}\,,
M. Faber\,\thanks{E--mail: faber@kph.tuwien.ac.at, Tel.:
+43--1--58801--14261, Fax: +43--1--58801--14299}\,, H.
Fuhrmann\,\thanks{E--mail: hermann.fuhrmann@oeaw.ac.at}\,,\\
V. A. Ivanova\,\thanks{E--mail: viola@kph.tuwien.ac.at, State
Polytechnic University, Department of Nuclear Physics, 195251
St. Petersburg, Russian Federation}\,, J. Marton\,\thanks{E--mail:
johann.marton@oeaw.ac.at}\,, N. I. Troitskaya\,\thanks{State
Polytechnic University, Department of Nuclear Physics, 195251
St. Petersburg, Russian Federation}~~, J. Zmeskal\,\thanks{E--mail:
johann.zmeskal@oeaw.ac.at}}

\date{\today}

\maketitle

\vspace{-0.5in}

\begin{center}
{\it Atominstitut der \"Osterreichischen Universit\"aten,
Arbeitsbereich Kernphysik und Nukleare Astrophysik, Technische
Universit\"at Wien, \\ Wiedner Hauptstrasse 8-10, A-1040 Wien,
\"Osterreich \\ and\\ Stefan Meyer Institut f\"ur subatomare Physik 
der \"Osterreichische Akademie der Wissenschaften,\\
Boltzmanngasse 3, A-1090, Wien, \"Osterreich} 
\end{center}

\begin{center}
\begin{abstract}
  In the model of low--energy $\bar{K}N$ interactions near threshold
  (EPJA {\bf 21}, 11 (2004); {\bf 25}, 79 (2005)) we calculate
  isospin--breaking corrections to the energy level displacement of
  the ground state of kaonic hydrogen, investigated by Mei\ss ner,
  Raha and Rusetsky (EPJC {\bf 35}, 349 (2004)) within the
  non--relativistic effective Lagrangian approach based on ChPT by
  Gasser and Leutwyler. Our results agree well with those by Mei\ss
  ner {\it et al.}. In addition we calculate the dispersive
  corrections, caused by the transition $K^-p \to \bar{K}^0n \to K^-p$
  with the $\bar{K}^0n$ pair on--mass shell. We show also how hypothesis
  on the dominant role of the $\bar{K}^0n$--cusp for the S--wave
  amplitude of low--energy $K^-p$ scattering near threshold, used by
  Mei\ss ner {\it et al.}, can be realized in our approach.  The
  result agrees fully with that by Mei\ss ner {\it et al.}.
\end{abstract} 


\end{center}

\newpage

\section{Introduction}
\setcounter{equation}{0}

Recently Gasser has outlined the contemporary experimental and
theoretical status of kaonic atoms in low--energy QCD \cite{JG04} and
resumed that the experimental data on the energy level displacement of
the ground state of kaonic hydrogen
\begin{eqnarray}\label{label1.1}
  - \,\epsilon^{(\exp)}_{1s} + i\,\frac{\Gamma^{(\exp)}_{1s}}{2} = 
(- 194 \pm 41) + i\,(125 \pm 59)\,{\rm eV},
\end{eqnarray} 
obtained by the DEAR Collaboration \cite{DEAR1}, have been understood
theoretically within the Effective Field Theory (the EFT) approaches
\cite{IV3,UM04} (see also \cite{IV4}). Nevertheless, as has been
pointed by Gasser \cite{JG04}, only more precise experimental data can
testify whether the contemporary theoretic approaches are able to
describe the complicated situation properly\,\footnote{We cite Gasser
  \cite{JG04}: ``{\bf Conclusions}:\,3.  The ground state of {\it
    kaonic hydrogen} has been investigated in a beautiful experiment
  at DEAR$^{[7]}$.  Data are available, the S--states of the atom are
  theoretically understood$^{[18,17]}$.\,4.  On the other hand, the
  theory of $\bar{K}p$ scattering leaves many questions open. More
  precise data will reveal whether present techniques are able to
  describe the complicated situation properly.'' References [17] and
  [18] are Refs.\cite{IV3} and \cite{UM04} of this paper.}.  

The theoretical value for the energy level displacement of the ground
state of kaonic hydrogen, calculated in \cite{IV3,IV4}, is equal to 
\begin{eqnarray}\label{label1.2}
  - \,\epsilon^{(0)}_{1s} + i\,\frac{\Gamma^{(0)}_{1s}}{2} &=& 
  2\,\alpha^3\mu^2 
  \,f^{K^-p}_0(0) =  421\,[{\cal R}e\,f^{K^-p}_0(0) + 
  i\,{\cal I}m\,f^{K^-p}_0(0)] = \nonumber\\
  &=& (- 203 \pm 15) + i\,(113 \pm 14)\,{\rm eV}, 
\end{eqnarray}
where $2\alpha^3\mu^2 = 421\,{\rm eV}/{\rm fm}$ with $\mu = m_K
m_p/(m_K + m_p) = 324\,{\rm MeV}$ is the reduced mass of the $K^-p$
pair, computed for $m_K = 494\,{\rm MeV}$ and $m_p = m_N = 940\,{\rm
  MeV}$, and $\alpha = 1/137.036$ is the fine--structure constant, 
then $f^{K^-p}_0(0)$ equal 
\begin{eqnarray}\label{label1.3}
f^{K^-p}_0(0) = {\cal R}e\,f^{K^-p}_0(0) + i\,{\cal
  I}m\,f^{K^-p}_0(0) =  (-\,0.482 \pm 0.034) + i\,(0.269 \pm
0.032)\,{\rm fm}
\end{eqnarray}
is the S--wave amplitude of $K^-p$ scattering at threshold \cite{IV3}.

The result Eq.(\ref{label1.2}) has been obtained within a quantum
field theoretic and relativistic covariant model of strong low--energy
$\bar{K}N$ interactions near threshold of $K^-p$ scattering. This is
an Effective Field Theory approach, formulated in terms of Effective
Chiral Lagrangians with a relativistic covariant quantization of
quantum fields. 

The S--wave amplitude of $K^-p$ scattering at threshold is computed to
leading order in chiral expansions of Chiral Perturbation Theory
(ChPT) by Gasser and Leutwyler \cite{JG83} (or to leading order
Effective Chiral Lagrangians \cite{MR96} (see also \cite{AI99})).

The imaginary part ${\cal I}m\,f^{K^-p}_0(0)$ of the S--wave amplitude
of $K^-p$ scattering at threshold is defined by inelastic channels
$K^-p \to \Sigma^-\pi^+$, $K^-p \to \Sigma^+\pi^-$, $K^-p \to
\Sigma^0\pi^0$ and $K^-p \to \Lambda^0\pi^0$, which are saturated by
the contributions of strange resonances $\Lambda(1405)$, 
$\Lambda(1800)$ and $\Sigma(1750)$ \cite{IV3}:
\begin{eqnarray}\label{label1.4}
 {\cal I}m\,f^{K^-p}_0(0) = {\cal I}m\,f^{K^-p}_0(0)_R.
\end{eqnarray}
In Gell--Mann's $SU(3)$ classification of hadrons, the
$\Lambda(1405)$--resonance is an $SU(3)$ singlet, whereas the
resonances $\Lambda(1800)$ and $\Sigma(1750)$ are components of an
$SU(3)$ octet \cite{PDG04}.

This allows to describe the experimental data on the cross sections
for inelastic reactions $K^-p \to Y\pi$, where $Y\pi = \Sigma^-\pi^+$,
$\Sigma^+\pi^-$, $\Sigma^0\pi^0$ and $\Lambda^0\pi^0$, at threshold of
the $K^-p$ pair \cite{DT71,RN78}
\begin{eqnarray}\label{label1.5}
\gamma &=& \frac{\sigma(K^-p \to \Sigma^-\pi^+)}{\sigma(K^-p \to
\Sigma^+\pi^-)} = 2.360 \pm 0.040,\nonumber\\ R_c &=&
\frac{\sigma(K^-p \to \Sigma^-\pi^+) + \sigma(K^-p \to
\Sigma^+\pi^-)}{\sigma(K^-p \to \Sigma^-\pi^+) + \sigma(K^-p \to
\Sigma^+\pi^-) + \sigma(K^-p \to \Sigma^0\pi^0) + \sigma(K^-p \to
\Lambda^0\pi^0)} = \nonumber\\ &=&0.664 \pm 0.011,\nonumber\\ R_n &=&
\frac{\sigma(K^-p \to \Lambda^0\pi^0)}{\sigma(K^-p \to \Sigma^0\pi^0)
+ \sigma(K^-p \to \Lambda^0\pi^0)} = 0.189 \pm 0.015
\end{eqnarray}
with an accuracy of about $6\,\%$ \cite{IV3}. According to Kaiser,
Siegel and Weise \cite{WW95}, any quantum field theoretic model of
low--energy $\bar{K}N$ interactions near threshold should reproduce
the experimental data Eq.(\ref{label1.5}). As a consequence of the
experimental data Eq.(\ref{label1.5}), we obtain also that the
$\Lambda(1800)$--resonance decouples from the $K^-p$ system at low
energies \cite{IV3}.

Due to the optical theorem the imaginary part ${\cal
  I}m\,f^{K^-p}_0(0)$ of the S--wave amplitude of $K^-p$ scattering
can be defined in terms of the parameters $\gamma$ and $R_c$. We get
\cite{IV3}:
\begin{eqnarray}\label{label1.6}
{\cal I}m\,f^{K^-p}_0(0) &=& |f(K^-p \to \Sigma^-\pi^+)|^2
k_{\Sigma^-\pi^+} + |f(K^-p \to \Sigma^+\pi^-)|^2
k_{\Sigma^+\pi^- }\nonumber\\
&+& |f(K^-p \to \Sigma^0\pi^0)|^2k_{\Sigma^0\pi^0} 
+ |f(K^-p \to \Lambda^0\pi^0)|^2
k_{\Lambda^0\pi^0}=\nonumber\\
&=&\frac{1}{R_c}\,\Big(1 +
\frac{1}{\gamma}\Big)\,|f(K^-p \to \Sigma^-\pi^+)|^2
k_{\Sigma^-\pi^+} = (0.269 \pm 0.032)\,{\rm fm},
\end{eqnarray}
where $k_{Y\pi}$ is a relative momentum of the $Y\pi$ system at
threshold of the $K^-p$ pair.

The S--wave amplitude $f(K^-p \to \Sigma^-\pi^+)$ of the inelastic
reaction $K^-p \to \Sigma^-\pi^+$ is equal to $f(K^-p \to
\Sigma^-\pi^+) = (0.379\pm 0.023)\,{\rm fm}$ \cite{IV3}. It is
computed with an accuracy of about $6\,\%$.  The main contribution to
$f(K^-p \to \Sigma^-\pi^+)$ comes from the $\Lambda(1405)$--resonance.
The contribution of the $\Sigma(1750)$--resonance makes up about
$2.7\,\%$ \cite{IV3}. This agrees well with a well--known assertion
that the $\Lambda(1405)$--resonance dominates in low--energy
interactions of the $K^-p$ pair in the S--wave state \cite{WW95}.

The real part ${\cal R}e\,f^{K^-p}_0(0)$  of the S--wave amplitude of $K^-p$
scattering at threshold, i.e. the S--wave scattering length of $K^-p$
scattering, 
\begin{eqnarray}\label{label1.7}
{\cal R}e\,f^{K^-p}_0(0) = {\cal R}e\,f^{K^-p}_0(0)_R
+{\cal R}e\,\tilde{f}^{K^-p}_0(0),
\end{eqnarray}
is defined by (i) the contribution of the strange baryon resonances
$\Lambda(1405)$ and $\Sigma(1750)$ in the s--channel of low--energy
elastic $K^-p$ scattering, (ii) the contribution of the exotic
four--quark (or $K\bar{K}$ molecules) scalar states $a_0(980)$ and
$f_0(980)$ in the $t$--channel of low--energy elastic $K^- p$
scattering and (iii) the contribution of low--energy interactions of
hadrons with non--exotic quark structures, i.e.  $q\bar{q}$ for mesons
and $qqq$ for baryons \cite{PDG04}.

We would like to notice that with an accuracy $6\,\%$ the contribution
of the strange resonances $\Lambda(1405)$ and $\Sigma(1750)$, which we
denote as ${\cal R}e\,f^{K^-p}_0(0)_R$, explains $32\,\%$ of the mean
value ${\cal R}e\,f^{K^-p}_0(0) = (-\,0.482 \pm 0.034) \,{\rm fm}$ of
the S--wave scattering length of $K^-p$ scattering. It is equal to
\cite{IV3}:
\begin{eqnarray}\label{label1.8}
{\cal R}e\,f^{K^-p}_0(0)_R &=&
\frac{1}{8\pi}\,\frac{\mu}{m_K}\,\Bigg[\frac{g^2_1}{\displaystyle
m_{\Lambda(1405)} - m_K - m_N} +
\frac{4}{9}\,\frac{g^2_2\,}{m_{\Sigma^0(1750)} - m_K -
m_N}\Bigg]=\nonumber\\ &=& (- 0.154 \pm 0.009)\,{\rm fm},
\end{eqnarray}
where the coupling constants $g_1 = 0.097$ and $g_2 = 1.123$ have been
calculated in \cite{IV3}.  The contribution of the
$\Sigma(1750)$--resonance makes up only $5.5\,\%$. This corroborates
again the dominance of the $\Lambda(1405)$--resonance in low--energy
interactions of the $K^-p$ pair in the S--wave state \cite{WW95}.

The contribution of the exotic scalar mesons $a_0(980)$ and $f_0(980)$
and non--exotic hadrons, calculated to leading order in chiral
expansion, is equal to \cite{IV3}:
\begin{eqnarray}\label{label1.9}
{\cal R}e\,\tilde{f}^{K^-p}_0(0) = 0.398 - 0.614\,\xi\,{\rm fm},
\end{eqnarray}
where the first term is defined by strong low--energy interactions of
non--exotic hadrons, whereas the second one is caused by the exotic
scalar mesons $a_0(980)$ and $f_0(980)$ coupled to the $K^-$--meson
and the proton.

The parameter $\xi$ is a low--energy constant (LEC), defining
low--energy $SNN$ interactions, where $S = a_0(980)$ and $f_0(980)$.

Using the hypothesis of the quark--hadron duality, formulated by
Shifman, Veinshtein and Zakharov within non--perturbative QCD
\cite{VZ79}, and the Effective quark model with chiral $U(3)\times
U(3)$ symmetry \cite{AI99}, we have calculated the parameter $\xi$
\cite{IV3}: $\xi = 1.2\pm 0.1$.

The validity of the application of the Effective quark model with
chiral $U(3)\times U(3)$ symmetry to the quantitative analysis of
strong low--energy $\bar{K}N$ interactions at threshold has been
corroborated by example of the calculation of the S--wave scattering
length of elastic $K^-n$ scattering in \cite{IV4} in connection with
the calculation of the energy level displacement of the ground state
of kaonic deuterium.

For the description of the contributions of the strange resonances
$\Lambda(1405)$ and $\Sigma(1750)$ and the exotic scalar mesons
$a_0(980)$ and $f_0(980)$ we have used Effective Chiral Lagrangians
with $\Lambda(1405)$, $\Sigma(1750)$, $a_0(980)$ and $f_0(980)$
treated as elementary particles, represented by local interpolating
fields.  Such a description of resonances and exotic mesons does not
contradict to ChPT by Gasser and Leutwyler \cite{JG83} and power
counting.

Indeed, as has been shown by Ecker {\it et al.} \cite{GE89}, the
exotic scalar mesons $a_0(980)$ and $f_0(980)$ can be described in
ChPT as elementary particle fields. The resonances $\Lambda(1405)$
and $\Sigma(1750)$ as well as the $\Delta(1232)$ can be also treated
in ChPT as elementary particle fields (baryon matter fields) without
contradiction to power counting \cite{MR96,UM01}. As has been pointed
out by Brown, Rho {\it et al.}  in \cite{MR96}, the inclusion of the
resonance $\Lambda(1405)$ as well as the resonance $\Sigma(1750)$ as
elementary particle fields (baryon matter fields) allows to compute
the amplitudes of $\bar{K}N$ scattering at threshold to leading order
Effective Chiral Lagrangians.

Recently, Ericson and Ivanov \cite{TE05} have shown that the treatment
of the $\Delta(1232)$--resonance as an elementary particle field gives
a possibility to compute the ChPT parameter $f_1$, describing in ChPT
isospin breaking phenomena in $\pi N$ scattering at threshold. As has
been shown in \cite{TE05}, the contribution of the
$\Delta(1232)$--resonance is necessary to saturate the value of $f_1$
in agreement with estimates obtained from different experimental data
and quark model approaches \cite{VL01}\,\footnote{We refer also on
  \cite{JG02}, where the constant $f_1$, computed within the quark
  model \cite{VL01}, has been used for the numerical estimates in
  ChPT.}.

The energy level shift and width $(\epsilon^{(0)}_{1s},
\Gamma^{(0)}_{1s})$ of the ground state of kaonic hydrogen
Eq.(\ref{label1.2}) have been calculated in \cite{IV3,IV4} at neglect
of corrections induced by isospin--breaking interactions.

A systematic analysis of corrections to the energy level displacement
of the ground state of kaonic hydrogen, caused by the
isospin--breaking interactions, has been recently carried out by
Mei\ss ner, Raha and Rusetsky \cite{UM04} within the non--relativistic
effective Lagrangian approach based on  ChPT by Gasser and
Leutwyler \cite{JG83}. The results are presented in terms of the
S--wave scattering lengths $a^I_0 = (a^0_0, a^1_0)$ of $\bar{K}N$
scattering with isospin $I = 0$ and $I = 1$, the mass difference $(m_d
- m_u)$ of current quark masses of the $u$ and $d$ quarks, which is a
quantity of order $(m_d - m_u) = O(\alpha)$ \cite{UM04}, and the
fine--structure constant $\alpha$.

The paper is organized as follows. In Section 2 we outline a
hypothesis on the dominant role of the $\bar{K}^0n$--cusp for the
S--wave amplitude of $K^-p$ scattering at threshold proposed by Dalitz
and Tuan in 1960 \cite{RD60,RD62} and reproduced by Mei\ss ner {\it et
  al.} within the non--relativistic effective Lagrangian approach
based on ChPT by Gasser and Leutwyler \cite{UM04}. In Section 3 we
calculate the dispersive corrections to the energy level displacement
of the ground state of kaonic hydrogen, caused by the contribution of
the transition $K^-p \to \bar{K}^0n \to K^-p$ with the $\bar{K}^0n$
pair on--mass shell.  We get $\delta^{Disp}_{EL} = (\delta^{Disp}_S,
\delta^{Disp}_W) = ((4.8 \pm 0.4)\,\%, (8.0\pm 1.0\,\%))$. We would
like to emphasize that such dispersive corrections from the
$\bar{K}^0n$ pair on--mass shell can exist only as corrections to the
energy level displacement of the ground state of kaonic hydrogen but
not to the S--wave amplitude of $K^-p$ scattering at threshold.
Therefore, these corrections could not be calculated by Mei\ss ner
{\it et al.}, who analysed the S--wave amplitude of $K^-p$ scattering
at threshold \cite{UM04}.  In Section 4 we calculate the corrections
to the energy level displacement of the ground state of kaonic
hydrogen, induced by isospin--breaking interactions. We get
$\delta^{ISB}_{EL} = (\delta^{ISB}_S, \delta^{ISB}_W) = (0.04\,\%,
-\,0.04\,\%)$.  These corrections can be attributed to $\delta {\cal
  T}_{KN} = O(\alpha)$, introduced by Mei\ss ner {\it et al.}
\cite{UM04}. Their values corroborate the smallness of $\delta {\cal
  T}_{KN} = O(\alpha)$ that agrees well with the assertion by Mei\ss
ner {\it et al.} \cite{UM04}.  In Section 5 we show how the hypothesis
on the $\bar{K}^0n$--cusp dominance for the S--wave scattering
amplitude of low--energy $K^-p$ scattering near threshold can be
realized in our approach. The result agrees fully with Mei\ss ner {\it
  et al.} \cite{UM04}. In the Conclusion we discuss the obtained
results. In Appendix we adduce the analytical calculation of the
momentum integrals in Eq.(\ref{label3.7}), defining the contribution
of the intermediate $\bar{K}^0n$ pair on--mass shell to the energy
level displacement of the ground state of kaonic hydrogen.

\section{On $K^-p$ scattering in S--wave state and $\bar{K}^0n$--cusp
  enhancement}
\setcounter{equation}{0}

For the numerical estimate of the energy level displacement of the
ground state of kaonic hydrogen Mei\ss ner {\it et al.} have proposed
no theoretical model for the calculation of the S--wave scattering
lengths $a^0_0$ and $a^1_0$ of $K^-N$ scattering.  Therefore, we
estimate the energy level displacement of the ground state of kaonic
hydrogen, obtained by Mei\ss ner {\it et al.}  \cite{UM04}, using the
S--wave scattering lengths $a^0_0$ and $a^1_0$, computed in our model
\cite{IV3,IV4}:
\begin{eqnarray}\label{label2.1}
a^0_0 &=& (-\,1.221 \pm 0.072) + i\,(0.537 \pm 0.064)\,{\rm fm},\nonumber\\ 
a^1_0 &=& (+\,0.258
\pm 0.024) + i\,(0.001 \pm 0.000) \,{\rm fm}.
\end{eqnarray}
For $m_d = m_u$ and the S--wave scattering lengths Eq.(\ref{label2.1})
the numerical value of the energy level displacement of the ground
state of kaonic hydrogen, obtained by Mei\ss ner {\it et al.}
\cite{UM04}, is equal to
\begin{eqnarray}\label{label2.2} 
  -\,\epsilon^{(\rm th)}_{1s} + i\,\frac{\Gamma^{(\rm th)}_{1s}}{2} = (- 266 \pm
  23) + i\,(177 \pm 21)\,{\rm eV}.
\end{eqnarray}
This agrees with both our theoretical result (\ref{label1.2}) and the
experimental data (\ref{label1.1}) by the DEAR Collaboration within
1.5 standard deviations for the shift and one standard deviation for
the width.  However, it does
not contradict also the experimental data by the KEK Collaboration
\cite{KEK}:
\begin{eqnarray}\label{label2.3} 
  -\,\epsilon^{(\exp)}_{1s} + i\,\frac{\Gamma^{(\exp)}_{1s}}{2} = 
(- 323 \pm 64) + i\,(204\pm 115)\,{\rm eV}.
\end{eqnarray}
Of course, one can try to solve an inverse problem.  Indeed, using the
experimental data by the DEAR Collaboration Eq.(\ref{label1.1}) or the
KEK Collaboration Eq.(\ref{label2.3}) and theoretical formula for the
energy level displacement of the ground state of kaonic hydrogen in
terms of $a^0_0$ and $a^1_0$ (see Eq.(\ref{label2.4})) to determine
the S--wave scattering lengths $a^0_0$ and $a^1_0$. Unfortunately, a
solution of this problem is rather ambiguous, since one needs to
define four parameters $({\cal R}e\,a^0_0, {\cal I}m\,a^0_0, {\cal
  R}e\,a^1_0, {\cal I}m\,a^1_0)$ in terms of two experimental values
$(\epsilon^{(\exp)}_{1s},\Gamma^{(\exp)}_{1s})$.

The main part of the S--wave amplitude of $K^-p$ scattering at
threshold is obtained in \cite{UM04} at the assumption of the dominant
role of the $\bar{K}^0n$--cusp, induced by the intermediate
$\bar{K}^0n$ state below threshold of $K^-p$ scattering, i.e. the
dominant role of the channel $K^-p \to \bar{K}^0n \to K^-p$.  The
hypothesis on the dominant role of the channel $K^-p \to \bar{K}^0n
\to K^-p$ for the S--wave amplitude of $K^-p$ scattering near
threshold has a long history. It has been proposed by Dalitz and Tuan
in 1960 \cite{RD60} (see also \cite{RD62}) and discussed by Deloff and
Law \cite{JL79}.

According to Dalitz and Tuan \cite{RD60,RD62} the S--wave amplitude
$\tilde{f}^{K^-p}_0(0)$ of $K^-p$ scattering at threshold, calculated
within the $K$--matrix approach in the zero--range approximation, is
defined by
\begin{eqnarray}\label{label2.4} 
  \tilde{f}^{K^-p}_0(0) = \frac{\displaystyle \frac{a^0_0 + a^1_0}{2}
+ q_0\,a^0_0a^1_0}{\displaystyle 1 + \Big(\frac{a^0_0 + a^1_0}{2}\Big)\,q_0},
\end{eqnarray}
where $q_0 = \sqrt{2\mu(m_{\bar{K}^0} - m_{K^-} + m_n - m_p)} =
58.4\,{\rm MeV} = O(\sqrt{\alpha})$ is a relative momentum of the
$\bar{K}^0n$ pair below threshold of $K^-p$ scattering, calculated for
$m_{\bar{K}^0} - m_{K^-} = 3.97\,{\rm MeV}$ and $m_n - m_p =
1.29\,{\rm MeV}$ \cite{PDG04}, $a^0_0$ and $a^1_0$ are complex S--wave
scattering lengths of $\bar{K}N$ scattering with isospin $I = 0$ and
$I = 1$.

The S--wave amplitude Eq.(\ref{label2.4}) has been reproduced by
Mei\ss ner {\it et al.} within the non--relativistic effective
Lagrangian approach based on ChPT by Gasser and Leutwyler (see Eq.(17)
of Ref.\cite{UM04}).

The energy level displacement of the ground state of kaonic hydrogen
is equal to $\tilde{f}^{K^-p}_0(0)$, given by Eq.(\ref{label2.4}),
multiplied by $2\alpha^3\mu^2 = 421\,{\rm eV}/{\rm fm}$ (see
Eq.(\ref{label1.2})). 

For the analysis of the contribution of the $\bar{K}^0n$--cusp it is
convenient to transcribe the S--wave amplitude Eq.(\ref{label2.4})
into the form: 
\begin{eqnarray}\label{label2.5} 
 \tilde{f}^{K^-p}_0(0) = a^{K^-p}_0 + \tilde{f}^{K^-p}_0(0)_{\rm cusp},
\end{eqnarray}
where $a^{K^-p}_0 = (a^0_0 + a^1_0)/2$ is the S--wave scattering
length of $K^-p$ scattering in the limit of isospin invariance, i.e.
at $q_0 = 0$, and $\tilde{f}^{K^-p}_0(0)_{\rm cusp}$ is defined by the
$\bar{K}^0n$--cusp for $q_0 \neq 0$:
\begin{eqnarray}\label{label2.6} 
  \tilde{f}^{K^-p}_0(0)_{\rm cusp} &=& -\,\Big(\frac{a^0_0 -  a^1_0}{2}\Big)^2
\frac{q_0}{\displaystyle 1 + \Big(\frac{a^0_0 + a^1_0}{2}\Big)\,q_0} = \nonumber\\
  &=& (-\,0.149 \pm 0.042) + 
  i\,(0.151 \pm 0.037)\,{\rm fm}.
\end{eqnarray}
The S--wave amplitude $\tilde{f}^{K^-p}_0(0)_{\rm cusp}$ has a
distinct form of the contribution of the inelastic channel $K^-p \to
\bar{K}^0n$ with the final--state $\bar{K}^0n$ interaction, calculated
in the zero--range approximation. The momentum $q_0$ in the numerator
of Eq.(\ref{label2.6}) is caused by the phase volume of the
$\bar{K}^0n$ pair in the S--wave state.  The numerical value of
$\tilde{f}^{K^-p}_0(0)_{\rm cusp}$ is calculated for the S--wave
scattering lengths Eq.(\ref{label2.1}). It is seen that the
corrections to the energy level displacement of the ground state of
kaonic hydrogen, caused by the $\bar{K}^0n$--cusp, make up about
$24\%$ and $36\%$ of the mean values of the energy level shift and
width given by Eq.(\ref{label2.2}), respectively. Such an enhancement
of the $K^-p$ interaction in the S--wave state at threshold agrees
with experimental data by the DEAR Collaboration within 1.5 standard
deviations for the shift and one standard deviation for the width.
There is no also contradiction to the experimental data by the KEK
Collaboration within the experimental error bars.

Hence, the improvement of the precision of the measurements of the
energy level displacement of the ground state of kaonic hydrogen by
the DEAR Collaboration and the KEK Collaboration should be important
for the confirmation of the hypothesis on the $\bar{K}^0n$--cusp
enhancement \cite{JG04}.

In Section 5 we show how the hypothesis on the dominant role of the
$\bar{K}^0n$--cusp for the S--wave scattering amplitude of low--energy
$K^-p$ scattering near threshold can be realized in our approach. This
testifies that our approach agrees fully with that by Mei\ss ner {\it
  et al.} \cite{UM04}. However, in this paper we do not follow the
hypothesis on the dominant role of the $\bar{K}^0n$--cusp in the
$K^-p$ interaction at threshold. 

Since the assumption of the dominant role of the $\bar{K}^0n$--cusp
for the S--wave amplitude of $K^-p$ scattering near threshold is a
dynamical hypothesis which should be confirmed or rejected by
experiments \cite{JG04}, the theoretical analysis of this hypothesis
goes beyond scope of this paper.

In this paper we focus our analysis on the confirmation of the
assertion by Mei\ss ner {\it et al.} that ``{\it $\ldots$, this
  piece of information\footnote{The $\bar{K}^0n$--cusp contribution to
    the S--wave amplitude of $K^-p$ scattering at threshold (see
    \cite{UM04}).}  should be still supplemented by the arguments in
  favour of the conjecture that the remaining corrections are small,
  i.e. $\delta {\cal T}_{KN} = O(\alpha)$, as done in the present
  paper \footnote{see Ref.\cite{UM04}.}.}''

We calculate the isospin--breaking corrections which can be attributed
to $\delta {\cal T}_{KN} = O(\alpha)$ (see \cite{UM04}), and compare
our results with those obtained by Mei\ss ner {\it et al.}
\cite{UM04}. In addition we compute the dispersive corrections to the
energy level displacement of the ground state of kaonic hydrogen,
induced by the intermediate $\bar{K}^0n$ state in the transition $K^-p
\to \bar{K}^0n \to K^-p$ with the $\bar{K}^0n$ pair on--mass shell.
Such corrections have been taken into account neither in our papers
\cite{IV3,IV4} nor in \cite{UM04}.

\section{Energy level displacement induced by the reaction $K^-p \to
  \bar{K}^0n \to K^-p$ with the $\bar{K}^0n$ pair on--mass shell}
\setcounter{equation}{0}

According to \cite{IV3}, the energy level displacement of the ground
state of kaonic hydrogen is defined by the S--wave amplitude of $K^-p$
scattering as follows
\begin{eqnarray}\label{label3.1}
  - \,\epsilon_{1s} + i\,\frac{\Gamma_{1s}}{2} &=&
  \frac{1}{4m_Km_p}\,\frac{1}{2}\sum_{\sigma_p = \pm 1/2}
\int\frac{d^3k}{(2\pi)^3}
  \int\frac{d^3q}{(2\pi)^3}\,
  \sqrt{\frac{m_K m_p}{E_K(\vec{k}\,)E_p(\vec{k}\,)}}\,
  \sqrt{\frac{m_K m_p}{E_K(\vec{q}\,)
      E_p(\vec{q}\,)}}\nonumber\\ &\times&\Phi^{\dagger}_{1s}(\vec{k}\,)\,
  M(K^-(\vec{q}\,)p(-\vec{q},\sigma_p) \to
  K^-(\vec{k}\,)p(-\vec{k},\sigma_p))\, \Phi_{1s}(\vec{q}\,),
\end{eqnarray}
where $M(K^-(\vec{q}\,)p(-\vec{q},\sigma_p) \to
K^-(\vec{k}\,)p(-\vec{k},\sigma_p))$ is the S--wave amplitude of
$K^-p$ scattering, $\Phi^{\dagger}_{1s}(\vec{k}\,)$ and
$\Phi_{1s}(\vec{q}\,)$ are the wave functions of the ground state of
kaonic hydrogen in the momentum representation normalized to unity: $
\Phi_{1s}(\vec{p}\,) = 8\sqrt{\pi a^3_B}/(1 + p^2 a^2_B)^2$; $a_B =
1/\alpha \mu = 83.6\,{\rm fm}$ is the Bohr radius.

By derivation of Eq.(\ref{label3.1}) \cite{IV3},
$M(K^-(\vec{q}\,)p(-\vec{q},\sigma_p) \to
K^-(\vec{k}\,)p(-\vec{k},\sigma_p))$ is an exact S--wave amplitude of
$K^-p$ scattering. It is defined also to arbitrary order in the
fine--structure constant expansion.

Due to the wave functions $\Phi^{\dagger}_{1s}(\vec{k}\,)$ and
$\Phi_{1s}(\vec{q}\,)$ the main contributions to the integrals over
$\vec{k}$ and $\vec{q}$ come from the regions of 3--momenta $k \sim
1/a_B$ and $q \sim 1/a_B$, where $1/a_B = 2.4\,{\rm MeV}$. Since the
integrand is concentrated around the momenta, which are much less than
masses of coupled particles $m_K \gg 1/a_B$ and $m_N \gg 1/a_B$, the
S--wave amplitude of $K^-p$ scattering can be defined in the
low--energy limit. The reduction of Eq.(\ref{label3.1}) to the
form of the well--known DGBTT
(Deser--Goldberger--Baumann--Thirring--Trueman) formula \cite{SD54}
has been discussed in \cite{IV3}.

The contribution of the intermediate $\bar{K}^0n$ state to the
energy level displacement of the ground state of kaonic hydrogen with
the $\bar{K}^0n$ pair on--mass shell can be defined as 
\cite{IV4,IV1,IV2}
\begin{eqnarray}\label{label3.2} 
  \hspace{-0.3in}&& - \,\delta \epsilon^{\bar{K}^0n}_{1s} + i\,
  \frac{\delta \Gamma^{\bar{K}^0n}_{1s}}{2} =\nonumber\\
  \hspace{-0.3in}&&=  \frac{1}{8m_Km_p}
  \int\frac{d^3k}{(2\pi)^3}
  \int\frac{d^3q}{(2\pi)^3}\,
  \sqrt{\frac{m_K m_p}{E_K(\vec{k}\,)E_p(\vec{k}\,)}}\,
  \sqrt{\frac{m_K m_p}{E_K(\vec{q}\,)
      E_p(\vec{q}\,)}}\Phi^{\dagger}_{1s}(\vec{k}\,)\,\Phi_{1s}(\vec{q}\,)
\nonumber\\
  \hspace{-0.3in}&&\times  \frac{1}{2}\sum_{\sigma_p = \pm 1/2}
  \sum_{\sigma_n = \pm 1/2}\int \frac{d^3Q}{(2\pi)^3 2E_{\bar{K}^0}(Q)
    2 E_n(Q)}\,
  \frac{1}{E_{\bar{K}^0}(Q) + E_n(Q) - E_K(k) - E_p(k) - i\,0}\, 
  \nonumber\\
  \hspace{-0.3in}&&\times
  M(K^-(\vec{q}\,)p(-\vec{q},\sigma_p) \to
  \bar{K}^0(\vec{Q}\,)n(-\vec{Q},\sigma_n))M(\bar{K}^0(\vec{Q}\,) 
  n(-\vec{Q},\sigma_n) \to
  K^-(\vec{k}\,)p(-\vec{k},\sigma_p))\nonumber\\
  \hspace{-0.3in} && + \frac{1}{8m_Km_p}
  \int\frac{d^3k}{(2\pi)^3}
  \int\frac{d^3q}{(2\pi)^3}\,
  \sqrt{\frac{m_K m_p}{E_K(\vec{k}\,)E_p(\vec{k}\,)}}\,
  \sqrt{\frac{m_K m_p}{E_K(\vec{q}\,)
      E_p(\vec{q}\,)}}\Phi^{\dagger}_{1s}(\vec{k}\,)\,\Phi_{1s}(\vec{q}\,)
  \nonumber\\
 \hspace{-0.3in} &&\times  \frac{1}{2}\sum_{\sigma_p = \pm 1/2}
  \sum_{\sigma_n = \pm 1/2}\int \frac{d^3Q}{(2\pi)^3 2E_{\bar{K}^0}(Q)
    2 E_n(Q)}\,
  \frac{1}{E_{\bar{K}^0}(Q) + E_n(Q) - E_K(q) - E_p(q) - i\,0}\, 
  \nonumber\\
 \hspace{-0.3in} &&\times
  M(K^-(\vec{q}\,)p(-\vec{q},\sigma_p) \to
  \bar{K}^0(\vec{Q}\,)n(-\vec{Q},\sigma_n))M(\bar{K}^0(\vec{Q}\,) 
  n(-\vec{Q},\sigma_n) \to
  K^-(\vec{k}\,)p(-\vec{k},\sigma_p)),\nonumber\\
 \hspace{-0.3in} &&
\end{eqnarray}
where $M(K^-(\vec{q}\,)p(-\vec{q},\sigma_p) \to
\bar{K}^0(\vec{Q}\,)n(-\vec{Q},\sigma_n))$ is the S--wave amplitude of
the inelastic reaction $K^-p \to \bar{K}^0n$.  To the imaginary part
of the S--wave amplitude of $K^-p$ scattering the inelastic channel
$K^-p \to \bar{K}^0n$ gives a contribution for relative momenta of the
$K^-p$ pair exceeding $q_0 = O(\sqrt{\alpha})$. Hence, the region of
the integration over $\vec{k}$ and $\vec{q}~$ is restricted from below
by $|\vec{k}\,|,|\vec{q}\,| \ge q_0$.

In the non--relativistic limit Eq.(\ref{label3.2}) reads
\begin{eqnarray}\label{label3.3}
 - \,\delta \epsilon^{\bar{K}^0n}_{1s} + i\,
  \frac{\delta \Gamma^{\bar{K}^0n}_{1s}}{2} &=& \frac{1}{\pi^2 \mu}
  \int\frac{d^3Q}{(2\pi)^3}\int^{\infty}_{q_0} 
  \frac{dk\,k^2\,\Phi_{1s}(k)\,
    f^*_{\bar{K}^0n}(Q)}{Q^2 + q^2_0 - k^2 - i\,0}
  \int^{\infty}_{q_0} dq\,q^2\, f_{\bar{K}^0n}(Q)\,\Phi_{1s}(q)\nonumber\\
  &+& \frac{1}{\pi^2 \mu}
  \int\frac{d^3Q}{(2\pi)^3}\int^{\infty}_{q_0} 
  dk\,k^2\,\Phi_{1s}(k)\,
  f^*_{\bar{K}^0n}(Q)
  \int^{\infty}_{q_0}\frac{dq\,q^2\, f_{\bar{K}^0n}(Q)\,
    \Phi_{1s}(q)}{Q^2 + q^2_0 - q^2 - i\,0},\nonumber\\ 
  && 
\end{eqnarray}
where $f_{\bar{K}^0n}(Q)$ is the S--wave amplitude of the reaction
$K^- p \to \bar{K}^0 n$ near threshold of the $\bar{K}^0n$ pair with
the final--state $\bar{K}^0n$ interaction \cite{IV4}, defined by
Feynman diagrams depicted in Fig.1. It is equal to
\begin{figure}
\centering \psfrag{K-}{$K^-$} 
\psfrag{p}{$p$}
\psfrag{n}{$n$}
\psfrag{=}{\large$=$}
\psfrag{+}{\large$+$}
\psfrag{K0}{$\bar{K}^0$}
\includegraphics[height=0.20\textheight]{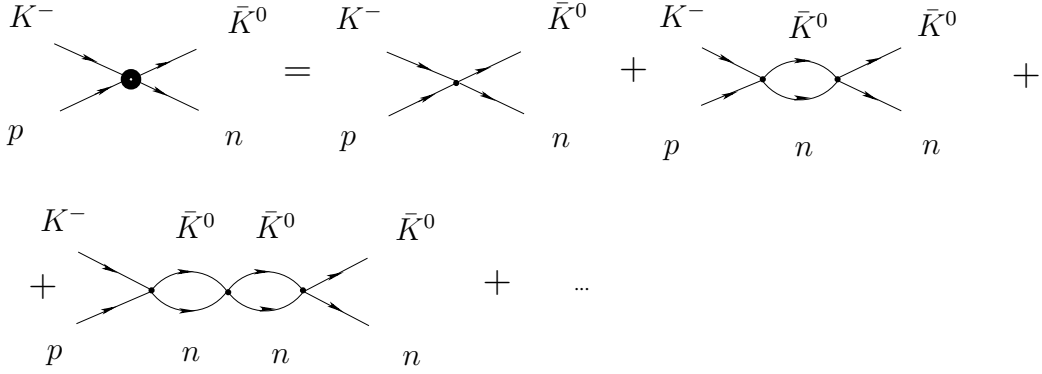}
\caption{Feynman diagrams of the amplitude of the reaction $K^-p \to
  \bar{K}^0 n$ taking into account the final--state $\bar{K}^0 n$
  interaction.}
\end{figure}
\begin{eqnarray}\label{label3.4}
  f_{\bar{K}^0n}(Q) = \frac{(a^1_0 - a^0_0)}{2}\,
  \frac{1}{\displaystyle 1 - i\,\Big(\frac{a^1_0 + a^0_0}{2}\Big)\,Q}.
\end{eqnarray}
The S--wave amplitude $f_{\bar{K}^0n}(Q)$ is taken in the zero--range
approximation. The contribution of the finite effective range
$r_{\bar{K}N}$ of low--energy $\bar{K}N$ scattering we discuss in the
end of this Section. 

Substituting Eq.(\ref{label3.4}) into Eq.(\ref{label3.3}) we obtain
\begin{eqnarray}\label{label3.5}
  \hspace{-0.3in}&& - \,\delta \epsilon^{\bar{K}^0n}_{1s} + i\,
  \frac{\delta \Gamma^{\bar{K}^0n}_{1s}}{2} = 
  \frac{(a^1_0 - a^0_0)^2}{8\pi^4 \mu}
  \int^{\infty}_{q_0}dk\,k^2\,\Phi_{1s}(k)\int^{\infty}_{q_0}
  dq\,q^2\,\Phi_{1s}(q)
  \nonumber\\ 
  \hspace{-0.3in}&&\times\int^{\infty}_0
  \frac{ dQ\,Q^2}{\displaystyle 1 + \Big(\frac{a^0_0 + a^1_0}{2}\Big)^2\,Q^2}\,
  \Big[\frac{1}{Q^2 + q^2_0 - k^2 - i\,0} 
  + \frac{1}{Q^2 + q^2_0 - q^2 - i\,0}\Big].
\end{eqnarray}
The integral over $Q$, calculated for $k \ge q_0$ and $q \ge
q_0$, reads
\begin{eqnarray}\label{label3.6}
  \hspace{-0.3in}&&\int^{\infty}_0
  \frac{ dQ\,Q^2}{\displaystyle 1 + \Big(\frac{a^0_0 + a^1_0}{2}\Big)^2\,Q^2}\,
  \Big[\frac{1}{Q^2 + q^2_0 - k^2 - i\,0} 
  + \frac{1}{Q^2 + q^2_0 - q^2 - i\,0}\Big] =\nonumber\\
  \hspace{-0.3in}&& = i\,\frac{\pi}{2}\,\Big[
  \frac{\sqrt{k^2 - q^2_0}}{\displaystyle 1 + 
    \Big(\frac{a^0_0 + a^1_0}{2}\Big)^2\,(k^2 - 
    q^2_0)} + 
  \frac{\sqrt{q^2 - q^2_0}}{\displaystyle 1 + 
    \Big(\frac{a^0_0 + a^1_0}{2}\Big)^2\,(q^2 - q^2_0)}\Big]\nonumber\\
  \hspace{-0.3in}&&+  \frac{\pi}{a^0_0 + a^1_0}\,
  \Big[\frac{1}{\displaystyle 1 + 
    \Big(\frac{a^0_0 + a^1_0}{2}\Big)^2\,(k^2 - q^2_0)}
  + \frac{1}{\displaystyle 1 + 
    \Big(\frac{a^0_0 + a^1_0}{2}\Big)^2\,(q^2 - q^2_0)}\Big].
\end{eqnarray}
The corrections $\delta \epsilon^{\bar{K}^0n}_{1s}$ and $\delta
\Gamma^{\bar{K}^0n}_{1s}$ are defined by
\begin{eqnarray}\label{label3.7}
  \hspace{-0.3in}\delta \epsilon^{\bar{K}^0n}_{1s} &=& - \frac{2\pi}{\mu}\,
  \Big(\frac{a^0_0 + a^1_0}{2}\Big)\,
  \Big(\frac{a^1_0 - a^0_0}{a^1_0 + a^0_0}\Big)^2\,\frac{1}{4\pi^4}
  \int^{\infty}_{q_0}\!\!\!\int^{\infty}_{q_0}\frac{dk\,dq\,k^2\,
    \Phi_{1s}(k)\,q^2\,
    \Phi_{1s}(q)}{\displaystyle  1 + 
    \Big(\frac{a^0_0 + a^1_0}{2}\Big)^2\,(q^2 - q^2_0)},\nonumber\\
  \hspace{-0.3in}\delta \Gamma^{\bar{K}^0n}_{1s} &=&
  \frac{(a^1_0 -a^0_0)^2}{8\pi^3 \mu}
  \int^{\infty}_{q_0}\!\!\!\int^{\infty}_{q_0}
  \frac{dk\,dq\,k^2\,
    \Phi_{1s}(k)\,q^2\,\sqrt{q^2 - q^2_0}\,
    \Phi_{1s}(q)}{\displaystyle  1 + 
    \Big(\frac{a^0_0 + a^1_0}{2}\Big)^2\,(q^2 - q^2_0)}.
\end{eqnarray}
The r.h.s. of Eq.(\ref{label3.7}) should be computed to leading order
in $q_0$--expansion, extracting from the energy level shift a
contribution independent on $q_0$. Indeed, such a contribution has
been already taken into account in \cite{IV3}. The momentum integrals
are calculated in Appendix. Keeping only the leading order
contributions in the $q_0$--expansion we get
\begin{eqnarray}\label{label3.8}
  \hspace{-0.3in}\delta^{Disp}_S &=& \frac{\delta 
    \epsilon^{\bar{K}^0n}_{1s}}{\epsilon^{(0)}_{1s}} = \frac{1}{4}\,
  (a^1_0 - a^0_0)^2\,q^2_0
  = (4.8 \pm 0.4)\,\%,\nonumber\\
  \hspace{-0.3in}\delta^{Disp}_W &=& \frac{\delta 
    \Gamma^{\bar{K}^0n}_{1s}}{\Gamma^{(0)}_{1s}} = 
  \frac{1}{2\pi}\,
  \frac{(a^1_0 -a^0_0)^2}{{\cal I}m\,f^{K^-p}_0(0)\,a_B}\,
  {\ell n}\Big[\frac{2 a_B}{|a^0_0 + a^1_0|}\Big] = (8.0 \pm 1.0)\,\%,
\end{eqnarray}
where ${\cal I}m\,f^{K^-p}_0(0) = (0.269 \pm 0.032)\,{\rm fm}$
\cite{IV3}. We have also taken into account the theoretical
uncertainty of the calculation of the shift of the energy level of the
ground state of kaonic hydrogen, which is equal to $\pm\,7\,\%$
\cite{IV3}.
 
The parameter $2/|a^0_0 + a^1_0| = (409 \pm 29)\,{\rm MeV}$, appearing
in the argument of the logarithm ${\ell n}(2a_B/|a^0_0 + a^1_0|)$ in
Eq.(\ref{label3.8}), has a meaning of a natural cut--off, introduced
by the final--state $\bar{K}^0n$ interaction.

The dispersive corrections Eq.(\ref{label3.8}) are caused by the
transition $K^-p \to \bar{K}^0n \to K^-p$ with the $\bar{K}^0n$ pair
on--mass shell. Hence, these corrections are only to the energy level
displacement of the ground state of kaonic hydrogen but not to the
S--wave amplitude of low--energy $K^-p$ scattering at threshold.  That
is why these corrections could not be calculated by Mei\ss ner {\it et
  al.}  \cite{UM04}, who analysed the isospin--breaking corrections to
the S--wave amplitude of $K^-p$ scattering at threshold.

Now let us discuss the contribution of the finite effective range of
low--energy $\bar{K}N$ scattering. The S--wave amplitude
$f_{\bar{K}^0n}(Q)$ in the finite effective--range approximation takes
the form \cite{MN79,TE88}
\begin{eqnarray}\label{label3.9}
  f_{\bar{K}^0n}(Q) = \frac{(a^1_0 - a^0_0)}{2}\,
  \frac{1}{\displaystyle 1 + \frac{1}{2}\,r_{\bar{K}N}\,
    \Big(\frac{a^1_0 + a^0_0}{2}\Big)\,Q^2 - i\,
    \Big(\frac{a^1_0 + a^0_0}{2}\Big)\,Q},
\end{eqnarray}
where $r_{\bar{K}N}$ is the effective range of low--energy $\bar{K}N$
scattering.  Unfortunately, there are no experimental data on the
value of the effective range $r_{\bar{K}N}$ \cite{MN79}. However,
without loss of generality one can assume that $r_{\bar{K}N}$, defined
by strong low--energy interactions, is smaller compared with the
S--wave scattering length, i.e. $r_{\bar{K}N} \ll 2/|a^0_0 + a^1_0|$.
In this case due to a fast convergence of the integral over $Q$ (see
Eq.(\ref{label3.6})) one can keep only the leading terms in the
effective range expansion. This leads to the replacement $(a^0_0 +
a^1_0) \to (a^0_0 + a^1_0 + r_{\bar{K}^0})$ in the integrand of
Eq.(\ref{label3.6}).

The correction to the energy level width $\delta^{Disp}_W$, amended by
the contribution of the finite value of the effective range
$r_{\bar{K}N}$, is equal now
\begin{eqnarray}\label{label3.10}
  \delta^{Disp}_W &=& 
  \frac{1}{4\pi}\,
  \frac{(a^1_0 -a^0_0)^2}{{\cal I}m\,f^{K^-p}_0(0)\,a_B}\,
  {\ell n}\Big[\frac{4a^2_B}{(a^0_0 + a^1_0 + r_{\bar{K}N})^2}\Big] = 
  \nonumber\\
  &=& \frac{1}{4\pi}\,
  \frac{(a^1_0 -a^0_0)^2}{{\cal I}m\,f^{K^-p}_0(0)\,a_B}\,\Big\{
  {\ell n}\Big[\frac{4 a^2_B}{(a^0_0 + a^1_0 )^2}\Big] + 
\frac{2\,r_{\bar{K}N}}{|a^0_0 + a^1_0|}\Big\}.
\end{eqnarray}
Thus, the contribution of the finite effective range of low--energy
$\bar{K}N$ scattering increases the value of the correction to the
energy level width.  However, due to the inequality $r_{\bar{K}N} \ll
2/|a^0_0 + a^1_0|$ the contribution of the finite effective range is
smaller compared with the mean value $(8.1 \pm 1.0)\,\%$, calculated
in the zero--effective range approximation.

For a finite effective range $r_{\bar{K}N} \ll 2/|a^0_0 + a^1_0|$ the
correction to the energy level shift $\delta^{Disp}_S$ acquires a
factor $(1 + 2r_{\bar{K}N}/(a^0_0 + a^1_0))$. This can change slightly
the value of $\delta^{Disp}_S$ relative to that, obtained in the
zero--range approximation.

\section{Energy level displacement induced by the QCD
  isospin--breaking interaction} 
\setcounter{equation}{0}

It is well--known that the contribution of the reaction $\pi^-p \to n
\gamma$, kinematically allowed at threshold of the $\pi^-p$ pair, to
the width of the energy level of the ground state of pionic hydrogen
makes up about $40\,\%$ \cite{PSI1}. In turn, the contribution of this
reaction to the energy level shift reduces to a level of a few
percents \cite{TE05}.

For kaonic hydrogen in the ground state there are two electromagnetic
reactions $K^-p \to \Lambda^0\gamma$ and $K^-p \to \Sigma^0\gamma$
allowed kinematically at threshold of the $K^-p$ pair in the S--wave
state. As has been shown in \cite{IV3}, the contribution of these
reactions to the width of the energy level of the ground state of
kaonic hydrogen makes up about $1\,\%$.  The correction to the energy
level shift is much less than $1\,\%$ \cite{TE05}.  Therefore, in this
Section we analyse the isospin--breaking corrections to the energy
level displacement of the ground state of kaonic hydrogen induced by
the QCD isospin--breaking interaction. Following Mei\ss ner {\it et
  al.} \cite{UM04} we take into account isospin--breaking corrections
of order $O(m_d - m_u) \sim O(\alpha)$ only.

According to Gasser and Leutwyler \cite{JG75,JG82}, the Lagrangian of
the QCD isospin--breaking interaction is equal to
\begin{eqnarray}\label{label4.1} 
{\cal L}^{\Delta I = 1}_{\rm \,QCD}(x) =
  \frac{1}{2}\,(m_d - m_u)\,[\bar{u}(x)u(x) - \bar{d}(x)d(x)],
\end{eqnarray}
where the current quark masses $m_u = 4\,{\rm MeV}$ and $m_d = 7\,{\rm
  MeV}$ are defined at the scale of the spontaneous breaking of chiral
symmetry $\Lambda_{\chi} \simeq 1\,{\rm GeV}$ \cite{JG75}.

In our model of strong low--energy $\bar{K}N$ interactions near
threshold \cite{IV3,IV4} the main contributions of the QCD
isospin--breaking interaction to the energy level displacement of the
ground state of kaonic hydrogen are defined by (i) the inelastic
channels $K^-p \to \Lambda(1405) \to \Lambda^0 (\eta(550) \to \pi^0)$
and $K^-p \to \Lambda(1405)\to (\Sigma^0 \to \Lambda^0)\pi^0$ to
${\cal I}m\,f^{K^-p}_0(0)$ (see Eq.(\ref{label1.6})) in terms of the
amplitudes of the isospin--breaking transitions $\eta(550) \to \pi^0$
and $\Sigma^0 \leftrightarrow \Lambda^0$, and (ii) the elastic channel
$K^-p \to \Lambda(1405) \to \bar{K}^0 n \to \Lambda(1405)\to K^-p$
with the $\bar{K}^0 n$ pair off--mass shell\,\footnote{The
  contribution of the intermediate $\bar{K}^0 n$ state with the
  $\bar{K}^0 n$ with the $\bar{K}^0n$ pair on--mass shell we have
  taken into account in Section 2.} to ${\cal R}e\,f^{K^-p}_0(0)_R$
(see Eq.(\ref{label1.8})) in terms of the mass differences
$m_{\bar{K}^0} - m_K$ and $m_n - m_p$.

The effective low--energy interactions of the
$\Lambda(1405)$--resonance, which we denote below as $\Lambda^0_1$,
with the $\Sigma^0\pi^0$, $K^-p$, $\bar{K}^0n$, and $\Lambda^0\eta$
pairs are defined by \cite{IV3}:
\begin{eqnarray}\label{label4.2}
{\cal L}_{\Lambda^0_1BP}(x) = g_1\bar{\Lambda}^0_1(x)
\Big(\Sigma^0(x)\pi^0(x) - 
p(x)K^-(x) + n(x)\bar{K}^0(x) +  \frac{1}{3}\Lambda^0(x)\eta(x)\Big) +
\ldots,
\end{eqnarray}
where $g_1 = 0.907$. The coupling constant $g_1 = 0.907$ has been
calculated in \cite{IV3} using the experimental data on the mass and
width of the $\Lambda(1405)$--resonance \cite{PDG04}.

\subsection{Imaginary part of the S--wave amplitude of $K^-p$
  scattering at threshold}

The S--wave amplitudes $f(K^-p \to \Sigma^0\pi^0)$ and $f(K^-p \to
\Lambda^0\pi^0)$ in Eq.(\ref{label1.6}), corrected by the
contributions of the QCD isospin--breaking interaction
Eq.(\ref{label4.1}), are equal to 
\begin{eqnarray}\label{label4.3}
  &&f(K^-p \to \Sigma^0\pi^0)= \frac{1}{4\pi}\,\frac{\mu}{m_K}\,\
  \sqrt{\frac{m_{\Sigma^0}}{m_N}}\,\Big[-\,\frac{g^2_1}{m_{\Lambda^0_1} - 
    m_K - m_N} \nonumber\\
  &&+ \,\frac{2}{3\sqrt{3}}\,\frac{g^2_2}{m_{\Sigma^0_2} - 
    m_K - m_N}\,\frac{1}{2}\,
  (m_d - m_u)\,\frac{\langle \Lambda^0|[\bar{u}(0)u(0) - \bar{d}(0)d(0)]
    |\Sigma^0\rangle}{m_{\Sigma^0} - m_{\Lambda^0}}\Big],\nonumber\\
  &&f(K^-p \to \Lambda^0\pi^0) =
  \frac{1}{4\pi}\,\frac{\mu}{~m_K}\,\sqrt{\frac{m_{\Lambda^0}}{m_N}}
  \Big[- \,\frac{2}{3\sqrt{3}}\,\frac{g^2_2}{m_{\Sigma^0_2} - 
    m_K - m_N}\nonumber\\
    && -\,\frac{g^2_1}{m_{\Lambda^0_1} - 
    m_K - m_N}\,\frac{1}{2}\,
  (m_d - m_u)\,\frac{\langle \Lambda^0|[\bar{u}(0)u(0) - \bar{d}(0)d(0)]
    |\Sigma^0\rangle}{m_{\Sigma^0} - m_{\Lambda^0}}\nonumber\\
    &&  -\,\frac{1}{3}\,\frac{g^2_1}{m_{\Lambda^0_1} - m_K - m_N}
\,\frac{1}{2}\,
  (m_d - m_u)\,\frac{\langle \eta|[\bar{u}(0)u(0) - \bar{d}(0)d(0)]
    |\pi^0\rangle}{m^2_{\eta} - m^2_{\pi}}\Big],
\end{eqnarray}
where $m_{\Sigma^0_2} = 1750\,{\rm MeV}$ is the mass of the
$\Sigma(1750)$ resonance and $g_2 = 1.123$ \cite{IV3}. The coupling
constant $g_2 = 1.123$ has been calculated in \cite{IV3} using the
experimental data on the width and mass of the
$\Sigma(1750)$--resonance \cite{PDG04}.

The isospin--breaking correction to the imaginary part of the S--wave
amplitude of $K^-p$ scattering at threshold is
\begin{eqnarray}\label{label4.4}
  \hspace{-0.3in}\delta^{ISB}_W &=& 
  \frac{\delta^{ISB}{\cal I}m\,f^{K^-p}_0(0)}{
    {\cal I}m\,f^{K^-p}_0(0)} = -\,\frac{4}{3\sqrt{3}}\,
  \frac{g^2_2}{g^2_1}\,\frac{m_{\Lambda^0_1} - m_K - m_N}{m_{\Sigma^0_2} - 
    m_K - m_N}\,\frac{|f(K^-p \to \Sigma^0\pi^0)|^2k_{\Sigma^0\pi^0}}{
    {\cal I}m\,f^{K^-p}_0(0)}\nonumber\\
  \hspace{-0.3in}&\times&\frac{1}{2}\,(m_d - m_u)\frac{\langle 
    \Sigma^0|[\bar{u}(0)u(0) - \bar{d}(0)d(0)]
    |\Lambda^0\rangle}{m_{\Sigma^0} - m_{\Lambda^0}} + 
  3\sqrt{3}\;\frac{g^2_1}{g^2_2}\,
  \frac{m_{\Sigma^0_2} - 
    m_K - m_N}{m_{\Lambda^0_1} - m_K - m_N}\nonumber\\
  \hspace{-0.3in}&\times&\frac{|f(K^-p \to 
    \Lambda^0\pi^0)|^2 
    k_{\Lambda^0\pi^0}}{{\cal I}m\,f^{K^-p}_0(0)}\, \frac{1}{2}\,
(m_d - m_u)\Big[\frac{\langle
    \Sigma^0|[\bar{u}(0)u(0) - \bar{d}(0)d(0)]
    |\Lambda^0\rangle}{m_{\Sigma^0} - m_{\Lambda^0}}\nonumber\\
  \hspace{-0.3in}&+& \frac{1}{3}\,
  \frac{\langle \eta|[\bar{u}(0)u(0) - \bar{d}(0)d(0)]
    |\pi^0\rangle}{m^2_{\eta} - m^2_{\pi}}\Big],
\end{eqnarray} 
where $k_{\Sigma^0\pi^0} = 183\,{\rm MeV}$ and $k_{\Lambda^0\pi^0} =
256\,{\rm MeV}$ are the relative momenta of the $\Sigma^0\pi^0$ and
$\Lambda^0\pi^0$ pairs at threshold of the initial $K^-p$ state. 

The effective Lagrangian of the transitions $\Sigma^0
\longleftrightarrow \Lambda^0$ and $\eta(550) \longleftrightarrow
\pi^0$, induced by the QCD isospin--breaking interaction
Eq.(\ref{label4.1}), can be written as \cite{JG82}
\begin{eqnarray}\label{label4.5}
  {\cal L}^{ISB}_{\rm eff}(x) &=& \theta^{ISB}_B\,(m_{\Sigma^0} - m_{\Lambda^0})\,
  (\bar{\Sigma}^0(x)\Lambda^0(x) + \bar{\Lambda}^0(x)\Sigma^0(x))\nonumber\\
&+& \theta^{ISB}_M\,
  (m^2_{\eta} - m^2_{\pi^0})\,\eta(x)\pi^0(x),
\end{eqnarray}
where $\theta^{ISB}_B$ and $ \theta^{ISB}_M$ are mixing angles defined by
\begin{eqnarray}\label{label4.6}
  \theta^{ISB}_B &=& \frac{1}{2}\,(m_d - m_u)\,\frac{\langle 
    \Sigma^0|[\bar{u}(0)u(0) - \bar{d}(0)d(0)]
    |\Lambda^0\rangle}{m_{\Sigma^0} - m_{\Lambda^0}}, \nonumber\\
 \theta^{ISB}_M &=& \frac{1}{2}\,(m_d - m_u)\,
  \frac{\langle \eta|[\bar{u}(0)u(0) - \bar{d}(0)d(0)]
    |\pi^0\rangle}{m^2_{\eta} - m^2_{\pi}}. 
\end{eqnarray}
Following Gasser and Leutwyler \cite{JG82}, we set $\theta^{ISB}_B =
\theta^{ISB}_M = \theta^{ISB}$:
\begin{eqnarray}\label{label4.7}
  \theta^{ISB} &=& \frac{1}{2}\,(m_d - m_u)\,\frac{\langle 
    \Sigma^0|[\bar{u}(0)u(0) - \bar{d}(0)d(0)]
    |\Lambda^0\rangle}{m_{\Sigma^0} - m_{\Lambda^0}} = \nonumber\\
  &=& \frac{1}{2}\,(m_d - m_u)\,
  \frac{\langle \eta|[\bar{u}(0)u(0) - \bar{d}(0)d(0)]
    |\pi^0\rangle}{m^2_{\eta} - m^2_{\pi}}. 
\end{eqnarray}
Assuming that the $\eta(550)$--meson is the eighth
component of the pseudoscalar octet \cite{JG83} (see also \cite{IV4}),
the matrix element $\langle \eta|[\bar{u}(0)u(0) -
\bar{d}(0)d(0)]|\pi^0\rangle$, calculated to leading order in
chiral expansion of ChPT \cite{JG83} and
Current Algebra \cite{SA68,HP73}, is equal to
\begin{eqnarray}\label{label4.8}
  \langle \eta|[\bar{u}(0)u(0) - \bar{d}(0)d(0)]|\pi^0\rangle = 
-\,\frac{2}{\sqrt{3}}\,\frac{1}{F^2_{\pi}}\,\langle \bar{u}(0)u(0)\rangle,
\end{eqnarray}
where $\langle \bar{u}(0)u(0)\rangle = \langle \bar{d}(0)d(0)\rangle =
-\,(255)^3\,{\rm MeV}^3$ is the quark condensate, calculated at
$\Lambda_{\chi} = 940\,{\rm MeV}$ \cite{AI99}, and $F_{\pi} =
92.4\,{\rm MeV}$, the PCAC constant \cite{PDG04}.

Substituting Eq.(\ref{label4.8}) into Eq.(\ref{label4.7}) we get
\begin{eqnarray}\label{label4.9}
  \theta^{ISB} = -\,\frac{1}{\sqrt{3}\,F^2_{\pi}}\,\frac{(m_d - m_u)\,
\langle \bar{u}(0)u(0)\rangle}{m^2_{\eta} - m^2_{\pi^0}} = 0.012. 
\end{eqnarray}
This result agrees well with that obtained by Gasser and Leutwyler
within ChPT \cite{JG82} (see Eq.(8.3) of Ref.\cite{JG82}):
\begin{eqnarray}\label{label4.10}
  \theta^{ISB} = \frac{1}{2}\,\arctan\Big(\frac{\sqrt{3}\,(m_d - m_u)}{2 m_s 
- m_d - m_u}\Big) = 0.010,
\end{eqnarray}
calculated for $m_u = 4\,{\rm MeV}$, $m_d = 7\,{\rm MeV}$ and $m_s =
135\,{\rm MeV}$ \cite{JG75,JG82}. 

The values of the current quark masses $m_u = 4\,{\rm MeV}$ and $m_d =
7\,{\rm MeV}$ and $m_s = 135\,{\rm MeV}$ have been used in the
Effective quark model with chiral $U(3)\times U(3)$ symmetry for the
analysis of strong low--energy interactions \cite{AI99}. According to
\cite{AI99}, the scale of spontaneous breaking of chiral symmetry is
equal to $\Lambda_{\chi} = 0.94\,{\rm GeV}$, that agrees well with
$\Lambda_{\chi} \simeq 1\,{\rm GeV}$ \cite{JG75}.

We would like to emphasize that in QCD the product $m_q \langle
\bar{q}q\rangle = m_q(\mu) \langle \bar{q}q\rangle(\mu)$ is an
observable renormalization group invariant quantity, independent on
the renormalization scale $\mu$ \cite{FY83}. Therefore, the product
$(m_d - m_u)\,\langle \bar{u}(0)u(0)\rangle$, calculated for the
normalization scale $\mu$ equal the scale of spontaneous breaking of
chiral symmetry $\mu = \Lambda_{\chi} \simeq 1\,{\rm GeV}$, i.e.
\[(m_d(\mu) - m_u(\mu))\,\langle \bar{u}(0)u(0)\rangle(\mu) =  
(m_d(\Lambda_{\chi}) - m_u(\Lambda_{\chi}))\,\langle
\bar{u}(0)u(0)\rangle(\Lambda_{\chi})= -\,(255\,{\rm MeV})^3,\]
is observable. 

Due to the relation Eq.(\ref{label4.7}) the r.h.s. of
Eq.(\ref{label4.4}) can be transcribed into the form
\begin{eqnarray}\label{label4.11}
  \hspace{-0.3in}&&\delta^{ISB}_W = 
\frac{\delta^{ISB}{\cal I}m f^{K^-p}_0(0)}{
    {\cal I}m\,f^{K^-p}_0(0)} = 4\sqrt{3}\,\theta^{ISB}\Big[ -\,\frac{1}{9}\;
  \frac{g^2_2}{g^2_1}\,\frac{m_{\Lambda^0_1} - m_K - m_N}{m_{\Sigma^0_2} - 
    m_K - m_N}\,\frac{|f(K^-p \to \Sigma^0\pi^0)|^2k_{\Sigma^0\pi^0}}{
    {\cal I}m\,f^{K^-p}_0(0)}\nonumber\\
  \hspace{-0.3in}&&+ \,\frac{g^2_1}{g^2_2}\;
  \frac{m_{\Sigma^0_2} - 
    m_K - m_N}{m_{\Lambda^0_1} - m_K - m_N}\,\frac{|f(K^-p \to 
    \Lambda^0\pi^0)|^2 
    k_{\Lambda^0\pi^0}}{{\cal I}m\,f^{K^-p}_0(0)}\Big] = -\,0.04\,\%.
\end{eqnarray} 
Hence, the correction, induced by the QCD isospin--breaking
interaction, to the width of the energy level of the ground state of
kaonic hydrogen is negligible in agreement with the assertion by
Mei\ss ner {\it et al.} \cite{UM04}.

\subsection{Real part of the S--wave amplitude of $K^-p$
  scattering at threshold}

The isospin--breaking correction to ${\cal R}e\,f^{K^-p}_0(0)$ is
equal to
\begin{eqnarray}\label{label4.12}
\delta^{ISB}_S =   \frac{\delta^{ISB}{\cal R}e\,f^{K^-p}_0(0)}{
    {\cal R}e\,f^{K^-p}_0(0)} &=& \frac{1}{{\cal R}e\,f^{K^-p}_0(0)}\,
\frac{1}{8\pi}\frac{\mu}{~m_K}\Big(\frac{g^2_1}{m_{\Lambda^0_1}
 - m_K - m_N}\Big)^2\nonumber\\
&&\times\,{\cal F.P.}\int \frac{d^4p}{(2\pi)^4i}\frac{1}{m^2_{\bar{K}^0} - p^2}\,
\frac{m_n}{m^2_n - p^2},
\end{eqnarray} 
where ${\cal R}e\,f^{K^-p}_0(0) = (-\,0.482 \pm 0.034)\,{\rm fm}$
\cite{IV3} and ${\cal F.P.}$ stands for the {\it finite part} of the
integral over $p$.  The correction $\delta^{ISB}_S$ is defined by the
contribution of the transition $K^-p \to \Lambda(1405) \to \bar{K}^0 n
\to \Lambda(1405)\to K^-p$ with the $\bar{K}^0 n$ pair off--mass shell
to ${\cal R}e\,f^{K^-p}_0(0)_R$ (see Eq.(\ref{label1.8})). 
 
Expanding the integrand in powers of $(m_{\bar{K}^0} - m_{K^-})$
and $(m_n - m_p)$ and keeping only the leading terms we get
\begin{eqnarray}\label{label4.13}
  &&{\cal F.P.}\int \frac{d^4p}{(2\pi)^4i}\,\frac{1}{m^2_{\bar{K}^0} - p^2}\,
  \frac{m_n}{m^2_n - p^2} = -\,(m^2_{\bar{K}^0} - m^2_{K^-})
  \int \frac{d^4p}{(2\pi)^4i}\,\frac{1}{(m^2_K - p^2)^2}\,
  \frac{m_N}{m^2_N - p^2}\nonumber\\
  && -\,2\,(m_n - m_p)
  \int \frac{d^4p}{(2\pi)^4i}\,\frac{m^2_N}{m^2_K - p^2}\,
  \frac{1}{(m^2_N - p^2)^2} = \frac{1}{8\pi^2}\,\Big(\frac{m_n -m_p}{m_N} - 
  \frac{m_{\bar{K}^0} - m_K}{m_K}\Big)\nonumber\\
  &&\times\,\frac{m^2_K m^3_N}{(m^2_N - m^2_K)^2}\,
  {\ell n}\Big(\frac{m^2_N}{m^2_K}\Big) +\frac{1}{8\pi^2}\,\Big( 
  \frac{m_{\bar{K}^0} - m_K}{m_K}\,\frac{m^2_K}{m^2_N} - 
  \frac{m_n -m_p}{m_N}\Big)\,\frac{m^3_N}{m^2_N - m^2_K}.
\end{eqnarray} 
Substituting Eq.(\ref{label4.13}) into Eq.(\ref{label4.12}) we obtain
\begin{eqnarray}\label{label4.14}
  \hspace{-0.3in}\delta^{ISB}_S &=& 
  \frac{\delta^{ISB}{\cal R}e\,f^{K^-p}_0(0)}{
    {\cal R}e\,f^{K^-p}_0(0)} = \frac{1}{{\cal R}e\,f^{K^-p}_0(0)}\,
  \frac{1}{64\pi^3}\frac{\mu}{~m_K}\Big(\frac{g^2_1}{m_{\Lambda^0_1}
    - m_K - m_N}\Big)^2\nonumber\\
  \hspace{-0.3in} &\times&\Big\{\Big(\frac{m_n -m_p}{m_N} - 
  \frac{m_{\bar{K}^0} - m_K}{m_K}\Big)\,\frac{m^2_K m^3_N}{(m^2_N - m^2_K)^2}\,
  {\ell n}\Big(\frac{m^2_N}{m^2_K}\Big)\nonumber\\
  \hspace{-0.3in}  &+&
  \Big(\frac{m_{\bar{K}^0} - m_K}{m_K}\,\frac{m^2_K}{m^2_N} - 
  \frac{m_n -m_p}{m_N}\Big)\,\frac{m^3_N}{m^2_N - m^2_K}\Big\} = 0.04\,\%.
\end{eqnarray} 
The isospin--breaking correction $\delta^{ISB}_S = 0.04\,\%$ is of
order $O(\alpha)$.

The isospin--breaking corrections, computed in Section 3, can be
attributed to $\delta {\cal T}_{KN} = O(\alpha)$ of Ref.\cite{UM04}.
We have shown that they are negligible small in agreement with the
assertion by Mei\ss ner {\it et al.}  \cite{UM04}.

\section{On $\bar{K}^0n$--cusp enhancement}
\setcounter{equation}{0}

In this Section we show how the hypothesis on the dominant role of the
$\bar{K}^0n$--cusp contribution to the S--wave amplitude of
low--energy $K^-p$ scattering can be incorporated in our approach.
For this aim it suffices to use Eq.(\ref{label3.1}) and to determine
the amplitude of $K^-p$ scattering in accordance with Dalitz and Tuan
for $m_{\bar{K}^0} - m_K \neq 0$ and $m_n - m_p \neq 0$
\cite{RD60,RD62}. Due to the wave functions of kaonic hydrogen in the
ground state, the S--wave amplitude of $K^-p$ scattering in
Eq.(\ref{label3.1}) can be determined near threshold of $K^-p$
scattering as \cite{MN79}:
\begin{eqnarray}\label{label5.1} 
  M(K^-(\vec{q}\,)p(-\vec{q},\sigma_p) \to
  K^-(\vec{k}\,)p(-\vec{k},\sigma_p)) = 8\pi\,(m_K + m_N)\,f^{K^-p}_0(\sqrt{kq}),
\end{eqnarray}
where $f^{K^-p}_0(p)$ is equal to \cite{RD60,RD62}:
\begin{eqnarray}\label{label5.2} 
  f^{K^-p}_0(p) = \frac{\displaystyle \frac{a^0_0 + a^1_0}{2}
    -i\,\sqrt{p^2 - q^2_0}\,a^0_0a^1_0}{\displaystyle 1 - i\,
    \Big(\frac{a^0_0 + a^1_0}{2}\Big)\,\sqrt{p^2 - q^2_0} - 
    \,i\,p\,\Big(\frac{a^0_0 + a^1_0}{2} -
    i\,\sqrt{p^2 - q^2_0}\,a^0_0\,a^1_0\Big)}.
\end{eqnarray}
Here $\sqrt{p^2 - q^2_0}$ is a relative momentum of the $\bar{K}^0n$
pair, tending to $iq_0$ at $p \to 0$, i.e.  $\sqrt{p^2 - q^2_0} \to
i\,q_0$ at $p \to 0$.

Substituting the S--wave amplitude Eq.(\ref{label5.1}) into
Eq.(\ref{label3.1}) and taking into account that due to the wave
functions of the ground state of kaonic hydrogen the main
contribution to the integrand comes from the regions of 3--momenta $k
\sim 1/a_B$ and $q \sim 1/a_B$, where $1/a_B = 2.4\,{\rm MeV}$, the
energy level displacement of the ground state of kaonic hydrogen is
determined by
\begin{eqnarray}\label{label5.3}
  - \,\epsilon^{\rm DT}_{1s} + i\,\frac{\Gamma^{\rm DT}_{1s}}{2} = 
 \frac{2\pi}{\mu}\,\Big\{\tilde{f}^{K^-p}_0(0)\,|\Psi_{1s}(0)|^2 + 
i\,\Big[\tilde{f}^{K^-p}_0(0)\Big]^2\Big|\int \frac{d^3k}{(2\pi)^3}\,
\sqrt{k}\,\Phi_{1s}(k)\Big|^2\Big\},
\end{eqnarray}
where $\tilde{f}^{K^-p}_0(0)$ is given by Eq.(\ref{label2.4}). As the
integral over $k$ is equal to
\begin{eqnarray}\label{label5.4}
  \int \frac{d^3k}{(2\pi)^3}\,\sqrt{k}\,\Phi_{1s}(k) = 
\frac{3}{\sqrt{2 a_B}}\,|\Psi_{1s}(0)|,
\end{eqnarray}
the r.h.s. of Eq.(\ref{label5.3}) reads
\begin{eqnarray}\label{label5.5}
  - \,\epsilon^{\rm DT}_{1s} + i\,\frac{\Gamma^{\rm DT}_{1s}}{2} =  
\frac{2\pi}{\mu}\,\Big\{\tilde{f}^{K^-p}_0(0) + i\,\frac{9}{2a_B}\,
\Big[\tilde{f}^{K^-p}_0(0)\Big]^2\Big\}\,|\Psi_{1s}(0)|^2.
\end{eqnarray}
Following Mei\ss ner {\it et al.} \cite{UM04} and introducing the
notation ${\cal T}^{(0)}_{KN} = 4\pi\,(1 +
m_K/m_N)\,\tilde{f}^{K^-p}_0(0)$, we transcribe the r.h.s. of
Eq.(\ref{label5.5}) into the form
\begin{eqnarray}\label{label5.6}
  - \,\epsilon^{\rm DT}_{1s} + i\,\frac{\Gamma^{\rm DT}_{1s}}{2} =  
  \frac{{\cal T}_{KN}}{2m_K}\,\,|\Psi_{1s}(0)|^2 = 
\frac{\alpha^3\mu^3}{2\pi m_K}\,{\cal T}_{KN},
\end{eqnarray}
where we have used $|\Psi_{1s}(0)|^2 = \alpha^3 \mu^3/\pi$ and ${\cal
  T}_{KN}$ is given by
\begin{eqnarray}\label{label5.7}
  {\cal T}_{KN} = {\cal T}^{(0)}_{KN} + i\,\frac{9}{4\pi}\,
\frac{\alpha \mu^2}{2m_K}\, \Big[{\cal T}^{(0)}_{KN}\Big]^2.
\end{eqnarray}
The expression Eq.(\ref{label5.7}) agrees well with Eq.(17) of 
Ref.\cite{UM04}
\begin{eqnarray}\label{label5.8}
  {\cal T}_{KN} = {\cal T}^{(0)}_{KN} + i\,
  \frac{\alpha \mu^2}{2m_K}\, \Big[{\cal T}^{(0)}_{KN}\Big]^2 + 
  \,\delta {\cal T}_{KN}.
\end{eqnarray}
The deviation of the second term in Eq.(\ref{label5.7}) from that in
Eq.(\ref{label5.8}) we define as $\delta {\cal T}_{KN}$, the relative
value of which is equal to
\begin{eqnarray}\label{label5.9}
  \Big|\frac{\delta {\cal T}_{KN}}{{\cal T}^{(0)}_{KN}}\Big| =  
2\pi\,\alpha\,\mu\,\Big(1 - \frac{9}{4\pi}\Big)\,|\tilde{f}^{K^-p}_0(0)| = 
1.4\,\%.
\end{eqnarray}
The smallness of this correction agrees well with the assertion by
Mei\ss ner {\it et al.} \cite{UM04}.

Of course, we understand that $\delta {\cal T}_{KN}$, given by
Eq.(\ref{label5.9}), does not describe the total set of corrections of
order $O(\alpha)$, which have been taken into account by Mei\ss ner
{\it et al.} \cite{UM04}. However, as has been pointed out by Mei\ss
ner {\it et al.}, these corrections can be neglected in comparison
with the main contribution defined by the first term ${\cal
  T}^{(0)}_{KN}$ in Eq.(\ref{label5.8}). This has been also
corroborated by our calculations in Section 3. We would like to note
that finally the second term in Eq.(\ref{label5.8}) has been also
included by Mei\ss ner {\it et al.} into $\delta {\cal T}_{KN}$ (see
Eq.(19) of Ref.\cite{UM04} and \cite{AR05}). Hence, it can be dropped
in comparison with ${\cal T}^{(0)}_{KN}$ as well.

\section{Conclusion}
\setcounter{equation}{0}

We have calculated the isospin--breaking corrections and the
dispersive corrections to the energy level displacement of the ground
state of kaonic hydrogen within the quantum field theoretic model of
low--energy $\bar{K}N$ interactions developed in \cite{IV3,IV4} (see
also \cite{IV1,IV2} and \cite{IV5,IV6}). The isospin--breaking
corrections are caused by the QCD isospin--breaking interaction,
whereas the dispersive corrections are defined by the contribution of
the transition $K^-p \to \bar{K}^on \to K^-p$ to the S--wave amplitude
of $K^-p$ scattering with the $\bar{K}^0n$ pair on--mass shell.

As has been shown in \cite{IV3} the contribution of the
electromagnetic reactions $K^-p \to \Lambda^0\gamma$ and $K^-p \to
\Sigma^0\gamma$, allowed kinematically at threshold of the $K^-p$
pair, to the energy level width of the ground state of kaonic hydrogen
makes up about $1\,\%$. The correction of these reactions to the
energy level shift is much less than $1\,\%$ \cite{TE05}.  Therefore,
in this paper we have calculated the isospin--breaking corrections
induced by the QCD isospin--breaking interaction Eq.(\ref{label4.1}).

We have got $\delta^{ISB}_{EL} = (\delta^{ISB}_S, \delta^{ISB}_W) =
(0.04\,\%,- \,0.04\,\%)$.  The contribution to the energy level width
$\delta^{ISB}_W = - \,0.04\,\%$ we have obtained in terms of the
corrections to the amplitudes of the inelastic channels $K^-p \to
\Sigma^0\pi^0$ and $K^-p \to \Lambda^0\pi^0$ with the isospin breaking
transitions $\Sigma^0 \to \Lambda^0$, $\Lambda^0 \to \Sigma^0$ and
$\eta(550) \to \pi^0$. This is a complete set of the QCD
isospin--breaking corrections at the tree--hadron level, i.e. to
leading order Effective Chiral Lagrangians.  The correction to the
energy level shift $\delta^{ISB}_S = 0.04\,\%$ is obtained in terms
the mass differences $(m_{\bar{K}^0} - m_K)$ and $(m_n - m_p)$, which
are defined by the electromagnetic and QCD isospin--breaking
interactions \cite{JG75,JG82}. It is induced by the transition $K^-p
\to \Lambda^0(1405) \to \bar{K}^0n \to \Lambda^0(1405) \to K^-p$ with
the $\bar{K}^0n$ pair off--mass shell. 

The isospin--breaking corrections $\delta^{ISB}_{EL} =
(\delta^{ISB}_S, \delta^{ISB}_W) = (0.04\,\%,- \,0.04\,\%)$ are of
order $O(\alpha)$. They can be attributed to the S--wave amplitude
$\delta {\cal T}_{KN} = O(\alpha)$, introduced by Mei\ss ner {\it et
  al.} \cite{UM04}. The smallness of these corrections
$\delta^{ISB}_{EL} = (\delta^{ISB}_S, \delta^{ISB}_W) = (0.04\,\%,-
\,0.04\,\%)$ agrees well with the assertion by Mei\ss ner {\it et al.}
that the contribution of $\delta {\cal T}_{KN} = O(\alpha)$ is
negligible small relative to the main part of the S--wave amplitude of
$K^-p$ scattering at threshold.

In addition to the isospin--breaking corrections $\delta^{ISB}_{EL}$
we have computed the dispersive corrections to the shift and width of
the energy level of the ground state of kaonic hydrogen, induced by
the transition $K^-p \to \bar{K}^0n \to K^-p$ with the $\bar{K}^0n$
pair on--mass shell.  They are equal to $\delta^{Disp}_S = (4.8 \pm 
0.4)\,\%$ and $\delta^{Disp}_W = (8.0\pm 1.0)\,\%$, respectively.
These corrections are defined only for the energy level displacement
through Eq.(\ref{label3.1}) and do not exist for the S--wave amplitude
of $K^-p$ scattering at threshold.  Therefore, they have not been
calculated by Mei\ss ner {\it et al.}, who have investigated the
S--wave amplitude of $K^-p$ scattering at threshold.  Such corrections
have not been also calculated in \cite{IV3}.

We have also shown that the hypothesis on the dominate role of the
$\bar{K}^0n$--cusp for the S--wave amplitude of $K^-p$ scattering near
threshold can be realized in our approach. The energy level
displacement of the ground state kaonic hydrogen as well as the
S--wave amplitude of $K^-p$ scattering at threshold are computed in
complete agreement with the results obtained by Mei\ss ner {\it et
  al.}  \cite{UM04}.

Hence, our approach to the description of low--energy dynamics of
strong low--energy $\bar{K}N$ interactions near threshold together
with Eq.(\ref{label3.1}) agrees well with general description of
strong low--energy interactions of hadrons within the
non--relativistic Effective Field Theory based on ChPT by Gasser and
Leutwyler \cite{UM04,JG83}.  Nevertheless, we would like to emphasize
that the hypothesis on the dominate role of the $\bar{K}^0n$--cusp is
a dynamical one, the correctness of which should be confirmed
experimentally through the improvement of the precision of
measurements \cite{JG04}. A theoretical discussion of this hypothesis
goes beyond scope of this paper.

\subsection{ Energy level shift and the $\sigma^{I =
    1}_{KN}(0)$--term}

In our approach, taking into account the dispersive corrections
Eq.(\ref{label3.8}), the energy level displacement of the ground state
of kaonic hydrogen is equal to
\begin{eqnarray}\label{label6.1}
  - \,\epsilon^{(\rm th)}_{1s} + i\,\frac{\Gamma^{(\rm th)}_{1s}}{2} = 
  (- 213 \pm 15) + i\,(123 \pm 14)\,{\rm eV}. 
\end{eqnarray}
The theoretical value of the energy level width fits well the mean
value of the experimental data Eq.(\ref{label1.1}). The discrepancy
between the theoretical value of the shift and the mean value of the
experimental data Eq.(\ref{label1.1}) can be improved by taking into
account the contribution of the $\sigma^{(I = 1)}_{KN}(0)$--term,
given by \cite{IV5}:
\begin{eqnarray}\label{label6.2}
   \delta \epsilon^{(\sigma)}_{1s} =  \frac{\alpha^3 \mu^3}{2\pi m_K 
F^2_K}\Big[\sigma^{(I = 1)}_{KN}(0) - 
  \frac{m^2_K}{4m_N}i\int d^4x\langle p(\vec{0},\sigma_p)
  |{\rm T}(J^{4+i5}_{50}(x)J^{4-i5}_{50}(0))|
  p(\vec{0},\sigma_p)\rangle\Big].
\end{eqnarray} 
Here $J^{4\pm i5}_{50}(x)$ are time--components of the axial--vector
hadronic currents $J^{4\pm i5}_{5\mu}(x)$, changing strangeness
$|\Delta S| = 1$, $F_K = 113\,{\rm MeV}$ is the PCAC constant of the
$K$--meson \cite{PDG04} and the $\sigma^{(I = 1)}_{KN}(0)$--term is
defined by \cite{ER72}--\cite{JG99}
\begin{eqnarray}\label{label6.3}
  \sigma^{(I = 1)}_{KN}(0) = \frac{m_u + m_s}{4m_N}\,
\langle p(\vec{0},\sigma_p)|\bar{u}(0)u(0) + \bar{s}(0)s(0)|p(\vec{0},
\sigma_p)\rangle,
\end{eqnarray}
where $u(0)$ and $s(0)$ are operators of the interpolating fields of
$u$ and $s$ current quarks \cite{FY83}.  

The correction $\delta \epsilon^{(\sigma)}_{1s}$ to the shift of the
energy level of the ground state of kaonic hydrogen, caused by the
$\sigma^{(I = 1)}_{KN}(0)$, is obtained from the S--wave amplitude of
$K^-p$ scattering, calculated to next--to--leading order in ChPT
expansion at the tree--hadron level \cite{IV5} and Current Algebra
\cite{SA68,HP73} (see also \cite{ER72}):
\begin{eqnarray}\label{label6.4}
 \hspace{-0.3in} &&4\pi\,\Big(1 + \frac{m_K}{m_N}\Big)\,f^{K^-p}_0(0) = 
\frac{m_K}{F^2_K} - 
  \frac{1}{F^2_K}\,\sigma^{(I = 1)}_{KN}(0)\nonumber\\
 \hspace{-0.3in}  &&\hspace{1in} + 
  \frac{m^2_K}{4m_N F^2_K}\,i\int d^4x\,\langle p(\vec{0},\sigma_p)
  |{\rm T}(J^{4+i5}_{50}(x)J^{4-i5}_{50}(0))|
  p(\vec{0},\sigma_p)\rangle.
\end{eqnarray} 
The contribution of the $\sigma^{(I = 1)}_{KN}(0)$--term, $-
\sigma^{(I = 1)}_{KN}(0)/F^2_K$, to the S--wave amplitude of $K^-p$
scattering in Eq.(\ref{label6.4}) has a standard structure
\cite{ER72}.

Since the first term, $m_K/F^2_K$ computed to leading order in chiral
expansion, has been already taken into account in \cite{IV3}, the
second term, $- \sigma^{(I = 1)}_{KN}(0)/F^2_K$, and the third one
define next--to--leading order corrections in chiral expansion to the
S--wave amplitude of $K^-p$ scattering at threshold. 

According to the DGBTT formula \cite{SD54} (see also \cite{IV3}), the
contribution of these terms to the energy level shift can be obtained
multiplying them by the factor $-  2\alpha^3\mu^2$ (see
Eq.(\ref{label1.2})).  Hence, the last two terms in Eq.(\ref{label6.4})
determine a next--to--leading order correction in chiral expansion,
which is a correction of order $O(\alpha^3)$ to the shift of the
energy level of the ground state of kaonic hydrogen.

Taking into account the contribution $\delta
\epsilon^{(\sigma)}_{1s}$, the total shift of the energy level of the
ground state of kaonic hydrogen is equal to
\begin{eqnarray}\label{label6.5}
  \hspace{-0.3in} \epsilon^{(\rm th)}_{1s} &=&  (213 \pm 15) +
  \frac{\alpha^3 \mu^3}{2\pi m_K 
    F^2_K}\,\sigma^{(I = 1)}_{KN}(0)\nonumber\\
  \hspace{-0.3in} &-& \frac{\alpha^3 \mu^3 m_K}{ 
    8\pi  F^2_K m_N}\,i\int d^4x\,\langle p(\vec{0},\sigma_p)
  |{\rm T}(J^{4+i5}_{50}(x)J^{4-i5}_{50}(0))|
  p(\vec{0},\sigma_p)\rangle.
\end{eqnarray}
The theoretical estimates of the value of $\sigma^{(I = 1)}_{KN}(0)$,
carried out within ChPT with a dimensional regularization of divergent
integrals, are converged around the number $\sigma^{(I = 1)}_{KN}(0) =
(200 \pm 50)\,{\rm MeV}$ \cite{VB93,BB99}.  Hence, the contribution of
$\sigma^{(I = 1)}_{KN}(0)$ to the energy level shift amounts to
\begin{eqnarray}\label{label6.6}
 \frac{\alpha^3 \mu^3}{2\pi m_K 
    F^2_K}\,\sigma^{(I = 1)}_{KN}(0)  =  (67\pm 17)\,{\rm eV}.
\end{eqnarray}
The total contribution to the shift of the energy level of the ground
state of kaonic hydrogen is given by
\begin{eqnarray}\label{label6.7}
   \epsilon^{(\rm th)}_{1s} =  (280 \pm 23) - 
  \frac{\alpha^3 \mu^3 m_K}{ 
    8\pi F^2_K m_N}\,i\int d^4x\,\langle p(\vec{0},\sigma_p)
  |{\rm T}(J^{4+i5}_{50}(x)J^{4-i5}_{50}(0))|
  p(\vec{0},\sigma_p)\rangle.
\end{eqnarray}
The theoretical analysis of the second term in
Eq.(\ref{label6.7}) is required for the correct
understanding of the contribution of the $\sigma^{I = 1}_{KN}(0)$--term
to the shift of the energy level of the ground state of kaonic hydrogen.

Of course, one can solve the inverse problem. Indeed, carrying
out a theoretical calculation of the term 
\begin{eqnarray}\label{label6.8}
  \frac{\alpha^3 \mu^3 m_K}{ 
    8\pi F^2_K m_N}\,i\int d^4x\,\langle p(\vec{0},\sigma_p)
  |{\rm T}(J^{4+i5}_{50}(x)J^{4-i5}_{50}(0))|
  p(\vec{0},\sigma_p)\rangle
\end{eqnarray}
in Eq.(\ref{label6.5}) and using the experimental data on the shift of
the energy level of the ground state of kaonic hydrogen, measured by
the DEAR Collaboration Eq.(\ref{label1.1}), one can extract the value
of the $\sigma^{(I = 1)}_{KN}(0)$--term. Such a value of the
$\sigma^{(I = 1)}_{KN}(0)$--term should be compared with the
theoretical predictions, obtained within the EFT approach based on
ChPT by Gasser and Leutwyler.  We are planning to perform this
analysis in our forthcoming publication.

\section*{Acknowledgement}

We are grateful to Torleif Ericson for helpful discussions and reading
the manuscript and to J\"urg Gasser for discussions of $\Sigma^0
\leftrightarrow \Lambda^0$ and $\eta(550) \leftrightarrow \pi^0$
mixing angles, caused the QCD isospin--breaking interaction. We thank
Akaki Rusetsky for the clarification of the results obtained in
\cite{UM04}.

\section*{Appendix. Calculation of momentum integrals}

In this Appendix we calculate the momentum integrals
Eq.(\ref{label3.7}), defining the dispersive corrections to the energy
level displacement of the ground state of kaonic hydrogen, caused by
the transition $K^-p \to \bar{K}^0n \to K^-p$ with the $\bar{K}^0n$
pair on--mass shell.

For the analysis of the correction $\delta \epsilon^{\bar{K}^0n}_{1s}$
to the energy level shift we have to calculate the double momentum
integral, which we denote as $f_S(q_0)$:

$$
f_S(q_0) = \int^{\infty}_{q_0}\!\!\!\int^{\infty}_{q_0}\frac{dk\,dq\,k^2\,
    \Phi_{1s}(k)\,q^2\,
    \Phi_{1s}(q)}{\displaystyle  1 + 
    \Big(\frac{a^0_0 + a^1_0}{2}\Big)^2\,(q^2 - q^2_0)}.\eqno({\rm A}.1)
$$

\noindent We have to expand $f_S(q_0)$ in powers of $q_0$ and to keep
only the first non--vanishing term dependent on $q_0$, since the
contribution of $f_S(0)$ has been already taken into account in
\cite{IV3}. The integral over $k$ is equal to

$$
\int^{\infty}_{q_0}dk\,k^2\, \Phi_{1s}(k) =
4\pi\,\Big[\arctan\Big(\frac{1}{x_0}\Big) +
\frac{x_0}{1 + x^2_0}\Big]\,|\Psi_{1s}(0)|,\eqno({\rm A}.2)
$$

\noindent where $x_0 = q_0a_B$ and $|\Psi_{1s}(0)| = 1/\sqrt{\pi a^3_B}$.

The result of the calculation of the integral over $q$ reads

$$
\int^{\infty}_{q_0}\frac{dq\,q^2\, \Phi_{1s}(q)}{\displaystyle 1 +
  \Big(\frac{a^0_0 + a^1_0}{2}\Big)^2\,(q^2 - q^2_0)} =
4\pi\,\Big\{\frac{1 + \lambda^2 - \lambda^2 x^2_0}{(1 - \lambda^2 -
  \lambda^2 x^2_0)^2} \arctan\Big(\frac{1}{x_0}\Big)
$$
$$
+ \frac{1}{1 -
  \lambda^2 - \lambda^2 x^2_0}\,\frac{x_0}{1 + x^2_0} - \frac{2
  \lambda \sqrt{1 - \lambda^2 x^2_0}}{(1 - \lambda^2 - \lambda^2
  x^2_0)^2}\, \arctan\Big(\frac{\sqrt{1 - \lambda^2 x^2_0}}{\lambda
  x_0}\Big)\Big\}\,|\Psi_{1s}(0)|,\eqno({\rm A}.3) 
$$

\noindent where we have denoted

$$
\lambda = \Big|\frac{a^0_0 + a^1_0}{2 a_B}\Big| = O(\alpha)\quad,
\quad \lambda x_0 = \Big|\frac{a^0_0 + a^1_0}{2}\Big|q_0 =
O(\sqrt{\alpha}).  \eqno({\rm A}.4) 
$$

\noindent The function $f_S(q_0)$ is defined by the product of the
expressions given by Eqs.({\rm A}.2) and ({\rm A}.3). Expanding
the function $f_S(q_0)$ in powers of $q_0$ we get

$$
f_S(q_0) = f_S(0) + 4\pi^4\,\Big(\frac{a^0_0 +
  a^1_0}{2}\Big)^2q^2_0\,|\Psi_{1s}(0)|^2 + \ldots\,. \eqno({\rm A}.5) 
$$

\noindent For the calculation of the contribution to the shift $\delta
\epsilon^{\bar{K}^0n}_{1s}$ we have to use only the term proportional
to $q^2_0$.

The correction $\delta \Gamma^{\bar{K}^0}_{1s}$ to the energy level
width, caused by the transition $K^-p \to \bar{K}^0n \to K^-p$ with
the $\bar{K}^0n$ pair on--mass shell, is defined by the double
momentum integral, which we denote as

$$
f_W(q_0) = \int^{\infty}_{q_0}\!\!\!\int^{\infty}_{q_0}\frac{dk\,dq\,k^2\,
    \Phi_{1s}(k)\,q^2\,\sqrt{q^2 - q^2_0}\,
    \Phi_{1s}(q)}{\displaystyle  1 + 
    \Big(\frac{a^0_0 + a^1_0}{2}\Big)^2\,(q^2 - q^2_0)}.\eqno({\rm A}.6)
$$

\noindent Since the integral over $k$ is given by Eq.({\rm A}.2), we
have to calculate the integral over $q$. The calculation runs as
follows

$$
\int^{\infty}_{q_0}\frac{dq\,q^2\,\sqrt{q^2 -
    q^2_0}\,a^4_B}{\displaystyle (1 + a^2_B q^2)^2\Big(1 +
  \Big(\frac{a^0_0 + a^1_0}{2}\Big)^2\,(q^2 - q^2_0)\Big)} = $$
$$
=
-\frac{1}{2}\frac{d}{da}\Bigg[\int^{\infty}_{x_0}\frac{dx\,x^2\,\sqrt{x^2
    - x^2_0}}{\displaystyle (a^2 + x^2)(1 + \lambda^2\,(x^2 - x^2_0))}
\Bigg]_{a = 1}= -\frac{1}{2}\frac{d}{da}\Bigg[- \frac{a^2}{1 -
  \lambda^2 a^2 - \lambda^2 x^2_0}
$$
$$
\times\int^{\infty}_{x_0}\frac{dx\,
  \sqrt{x^2 - x^2_0}}{a^2 + x^2} + \frac{1 - \lambda^2 x^2_0}{1 -
  \lambda^2 a^2 - \lambda^2
  x^2_0}\int^{\infty}_{x_0}\frac{dx\,\sqrt{x^2 - x^2_0}}{1 +
  \lambda^2\,(x^2 - x^2_0)}\Bigg]_{a = 1}.\eqno({\rm A}.7) 
$$

\noindent Then, we make a change of variables $x = x_0\cosh\varphi$.
This gives

$$
\int^{\infty}_{q_0}\frac{dq\,q^2\,\sqrt{q^2 -
    q^2_0}\,a^4_B}{\displaystyle (1 + a^2_B q^2)^2\Big(1 +
  \Big(\frac{a^0_0 + a^1_0}{2}\Big)^2\,(q^2 - q^2_0)\Big)} = 
$$
$$
= \frac{1}{2}\frac{\sqrt{1 - \lambda^2 x^2_0}}{(1 -
  \lambda^2 - \lambda^2 x^2_0)^2}\,{\ell n}\Bigg(\frac{1 + \sqrt{1 -
    \lambda^2x^2_0}}{1 - \sqrt{1 - \lambda^2x^2_0}}\Bigg) - \frac{1}{2}\,
\frac{1}{1 -
  \lambda^2 - \lambda^2 x^2_0}
$$
$$
- \frac{1}{4}\,\frac{2 + (1 - \lambda^2 - \lambda^2 x^2_0)\,x^2_0}{(1 - \lambda^2 - \lambda^2 x^2_0)^2 \sqrt{1 + x^2_0}}\,
{\ell n}\Bigg(\frac{\sqrt{1 +
    x^2_0} + 1}{\sqrt{1 + x^2_0} - 1}\Bigg). \eqno({\rm A}.8) 
$$

\noindent To leading order in $q_0$--expansion the integral over $q$
reads

$$
\int^{\infty}_{q_0}\frac{dq\,q^2\,\sqrt{q^2 -
    q^2_0}\,a^4_B}{\displaystyle (1 + a^2_B q^2)^2\Big(1 +
  \Big(\frac{a^0_0 + a^1_0}{2}\Big)^2\,(q^2 - q^2_0)\Big)} =
\Big(1 + 2 \lambda^2 + \frac{3}{2}\,\lambda^2
x^2_0\Big){\ell n}\Big(\frac{1}{\lambda}\Big) + \ldots\,.
\eqno({\rm A}.9) 
$$

\noindent Hence, the double momentum integral $f_W(q_0)$,
calculated to leading order in $q_0$--expansion, is equal to 

$$
f_W(q_0) = \frac{16\pi^3}{a_B}\,{\ell n}\Big[\frac{2
  a_B}{|a^0_0 + a^1_0|} \Big]\,|\Psi_{1s}(0)|^2.\eqno({\rm A}.10)
$$

\noindent The function $f_W(q_0)$, given by Eq.({\rm A}.10), defines
the correction $\delta \Gamma^{\bar{K}^0n}_{1s}$ to the width of the
energy level of the ground state of kaonic hydrogen
Eq.(\ref{label3.7}), caused by the transition $K^-p \to \bar{K}^0n \to
K^-p$ with the $\bar{K}^0n$ pair on--mass shell.

\end{document}